\documentclass[aps,pra,reprint,
twocolumn,amsmath,superscriptaddress,amssymb,showpacs,nofootinbib]{revtex4-1}
\usepackage{amsmath,braket,amssymb}
\usepackage[final]{graphicx}
\usepackage{subfigure}
\usepackage[english]{babel}
\usepackage[utf8]{inputenc}
\usepackage{mathtools}
\usepackage{empheq}
\usepackage{tensor}
\usepackage{cancel}

\usepackage{simplewick}
\usepackage{xfrac}

\usepackage[usenames,dvipsnames]{color}

\usepackage{enumitem}
\usepackage{slashed}
\usepackage{graphicx}
\usepackage{xcolor}
\usepackage{tikz}

\usepackage{bm}
\usepackage[colorlinks=true,citecolor=blue,linkcolor=blue,urlcolor=blue]{hyperref}
\usepackage{dsfont}
\usepackage{changepage}
\usepackage{array}
\usepackage{soul}
\usepackage[capitalise]{cleveref}
\crefname{section}{Sec.}{Secs.}
\Crefname{section}{Sec.}{Secs.}
\usepackage{xifthen}
\usepackage{xargs}

\usepackage{bm}
\renewcommand{\vec}[1]{\boldsymbol{\mathbf{#1}}}

\graphicspath{{./}{./figures/}}

\newcommand{\bit}{\begin{itemize}}
\newcommand{\eit}{\end{itemize}}

\newcommand{\f}{\frac}
\renewcommand{\>}{\right\rangle}
\newcommand{\<}{\left\langle}
\newcommand{\ba}{\begin{align}}
\newcommand{\ea}{\end{align}}
\newcommand{\be}{\begin{equation}}
\newcommand{\ee}{\end{equation}}
\newcommand{\bi}{\begin{itemize}}
\newcommand{\ei}{\end{itemize}}
\newcommand{\lf}{\left(}
\newcommand{\ri}{\right)}
\newcommand{\dd}{\mathrm{d}}

\newcommand{\Tr}{\operatorname{Tr}}

\DeclareMathAlphabet{\mymathbb}{U}{BOONDOX-ds}{m}{n}

\renewcommand{\log}{\ln}

\usepackage{comment}

\newcommand\wye[1][]{%
    \tikz\draw[thick, line cap=round,x=1ex,y=1ex,#1]
    (0,0) -- ++(90:1)
    (0,0) -- ++(-30:1)
    (0,0) -- ++(-150:1);
}

\usepackage{braket}
\usepackage[normalem]{ulem}

\newcommand{\nocontentsline}[3]{}
\newcommand{\tocless}[2]{\vspace{2em}\bgroup\let\addcontentsline=\nocontentsline#1{#2}\egroup}

\begin{document}

\title{Heisenberg spin chain with random-sign couplings}
\author{Michele Fava}
\affiliation{Philippe Meyer Institute, Physics Department, \'{E}cole Normale Sup\'{e}rieure (ENS), Universit\'{e} PSL, 24 rue Lhomond, F-75231 Paris, France}
\author{Jesper Lykke Jacobsen}
\affiliation{Laboratoire de Physique de l’\'Ecole Normale Sup\'erieure, CNRS, ENS \& Universit\'e PSL, Sorbonne Universit\'e, Universit\'e Paris Cit\'e, 75005 Paris, France}
\affiliation{Sorbonne Universit\'e, \'Ecole Normale Sup\'erieure, CNRS, Laboratoire de Physique (LPENS), F-75005 Paris, France}
\author{Adam Nahum}
\affiliation{Laboratoire de Physique de l’\'Ecole Normale Sup\'erieure, CNRS, ENS \& Universit\'e PSL, Sorbonne Universit\'e, Universit\'e Paris Cit\'e, 75005 Paris, France}

\date{\today}

\begin{abstract}
We study the 1D quantum Heisenberg chain with randomly ferromagnetic or antiferromagnetic couplings (a model previously studied by approximate strong-disorder RG). We find that, at least for sufficiently large spin $S$, the ground state has ``spin glass'' order. The spin waves on top of this state have the dynamical exponent ${z=3/2}$, intermediate between the values $z=1$ of the antiferromagnet and ${z=2}$ of the ferromagnet. DMRG simulations are in good agreement with the analytical results for spins ${S=1}$ and ${S=3/2}$. The case ${S=1/2}$ shows large finite size effects: we suggest that this case is also ordered, but with a small ordered moment.
\end{abstract}

\maketitle

\paragraph*{Significance statement.}
A class of magnetic alloys can be modeled by $\mathrm{SU}(2)$-invariant quantum spin chains in which each bond is randomly ferromagnetic or antiferromagnetic. Unlike other simple spin chains, the nature of the ground states of these chains has not been resolved (standard renormalization group approaches to random chains are not exact when ferromagnetic bonds are present). Contrary to the common assumption that one-dimensional systems with continuous symmetry (and a non-conserved order parameter) cannot spontaneously break symmetry, we show that the ground states can exhibit ``spin-glass'' order, together with unusual spin-wave fluctuations. The dynamical critical exponent $z$, which links fluctuation wavelength $\lambda$ and frequency $f$ as ${f\sim\lambda^{-z}}$, assumes a fractional value of $3/2$, halfway between the values for the purely ferromagnetic and antiferromagnetic chains.

\tocless\section{Introduction}
We analyze the low-energy physics of disordered spin chains of the form~\cite{PhysRevLett.73.2622,PhysRevB.52.15930,nguyen1996design,PhysRevLett.75.4302,sherrington1979long} 
\begin{equation}
    \label{eq:basicH}
    \mathcal{H} = - {\sum}_j\mathcal{J}_j \, \vec S_j \cdot \vec S_{j+1},
\end{equation}
where the signs of the couplings $\mathcal{J}_j$ are chosen randomly and independently. 
For definiteness we take ${\mathcal{J}_j = \pm \mathcal{J}}$ with equal probability. The size $S$ of the spins may be arbitrary, with $S=1/2$ in the minimal case.

A version of this model in which the magnitude of $\mathcal{J}_j$ is also random was studied in Refs.~\cite{PhysRevLett.75.4302,PhysRevB.55.12578,PhysRevB.60.12116} using the strong-disorder renormalization group (RG), 
in which the strongest bond in the system is decimated in each RG step~\cite{igloi2005strong}: 
it  was argued that  the RG process involves the formation of effective spins of successively larger magnitudes on increasing scales, leading to a correlated disordered state. Aspects of this picture were tested in quantum Monte Carlo~\cite{PhysRevB.60.3388, Ammon1999}.
However,
unlike the case of the antiferromagnetic spin-1/2 chain~\cite{ma1979random,PhysRevB.50.3799},
the strong-disorder RG does not become exact at large scales for models with both signs of $\mathcal{J}_j$~\cite{PhysRevLett.75.4302,PhysRevB.55.12578}. 
As a result, the universal long-wavelength behavior of this class of models has not been clarified.
Here we resolve the question for models like that defined above, where the dominant randomness is the sign-randomness of ${\mathcal{J}_j}$ (see below).
We argue that the ground states of such chains are, in fact, ordered (with adjacent moments being aligned or antialigned according to the sign of their coupling, as in a ``Mattis spin glass''~\cite{sherrington1979long,mattis1976solvable}), and thus show spontaneous (continuous) symmetry breaking, despite being in one dimension. The low-energy theory is nontrivial and is determined by spin waves with a spatially random stiffness matrix.

We begin by analyzing the quadratic Lagrangian for the spin waves, obtained via the large-$S$ expansion. 
The most basic universal quantity is the dynamical exponent, controlling the typical energy scale ${\Delta E \sim \ell^{-z}}$ for excitations on scale $\ell$. In fact, this is equivalent to the dynamical exponent for the purely classical chain, which was computed earlier using a transfer matrix method in Ref.~\cite{PhysRevB.38.4980} (see also Refs.~\cite{PhysRevB.40.4947, PhysRevB.40.4638, PhysRevB.56.2320, PhysRevB.57.8269})
and explained heuristically by  analogy with the ferrimagnet in Ref.~\cite{gurarie2003bosonic}. 
Consistent with Refs.~\cite{PhysRevB.38.4980,gurarie2003bosonic} we find (by a different method)  that ${z=3/2}$. We also determine the scaling dimensions of the basic operators in the quantum theory, giving the power-law decay of correlation functions in the ground state and confirming that long-range order is stable.
These analytic results are compared with numerical results for the quadratic spin-wave theory.

Next, we perform numerical density matrix renormalization group (DMRG) simulations of the ``full'' quantum problem defined by Eq.~\ref{eq:basicH} at finite spin~$S$, for ${S=1/2,\, 1,\, 3/2}$.
In order to test the theory, we look at finite-size estimates of the order parameter, and at the energy gap and correlation functions.

DMRG simulations for ${S=1}$ and ${S=3/2}$ show clear evidence of long range order. For ${S=1}$, where we access larger sizes, we also see reasonable agreement with the expected universal exponent for the gap. For ${S=1/2}$ we also give  evidence that the chain is long-range-ordered,  with an ordered moment that is strongly reduced by fluctuations.
However, at the available sizes, the $S=1/2$ data does not fit well to the universal exponents expected from linear spin wave theory. The simplest hypothesis is that this mismatch is due to finite size effects, which are  large for $S=1/2$.

The scaling arguments here extend to more general probability distributions for the bonds~$\mathcal{J}_j$, including the case where ferromagnetic and antiferromagnetic couplings have unequal probabilities.
Randomness in the amplitudes  $|\mathcal{J}_j|$ can also be included. 
However, if the probability distribution for $|\mathcal{J}_j|$ has a heavy enough tail as ${|\mathcal{J}_j|\to 0}$, then rare weak bonds become important and may change the scaling behavior. 
Within spin-wave theory, we find that long-range order in fact survives in this regime, with a modified dynamical exponent.

\tocless\section{Spin wave Lagrangians}

Let us assume open boundary conditions, and write the random couplings as 
\begin{align}
\mathcal{J}_j & = JS^{-1} \times c_j c_{j+1}, 
&
c_j = \pm 1.
\end{align}
In the classical limit, ${S\to\infty}$, the ground states have long-range order in the ``staggered'' spin 
${\vec N_j \equiv c_j \vec S_j}$.
Quantum fluctuations on top of such an ordered state can be studied using the coherent states path integral~\cite{altland2010condensed}. 
We will argue below that the ground states at finite $S$ also have long range order in the staggered spin.

It is useful to note that the spin $S_\text{tot}$ of the ground-state multiplet is determined rigorously by the sign structure of the Hamiltonian~\cite{lieb1962ordering}: if $n_A$ and $n_B$ are the numbers of sites with $c_j=1$ and $c_j=-1$ respectively, then ${S_\text{tot} = S |n_A-n_B|}$, and is therefore of order $S\sqrt{L}$ for typical disorder realizations.
(This scaling is also respected in the strong disorder RG.)

In the coherent states approach each spin $\vec S_j$ is represented by a field $S \vec n_j(t)$, 
and the imaginary-time Lagrangian~is (with ${\hbar=1}$)
 \begin{equation}\label{eq:initialaction}
\mathcal{L}= 
- i S  \sum_{j=1}^L  (1-z_j) \dot \theta_j
- 
S  \sum_{j=1}^{L-1}  J c_j c_{j+1} \vec n_j \cdot \vec n_{j+1}.
\end{equation}
The first term is the Berry phase~\cite{altland2010condensed}, written
in the parameterization 
${\vec n = \lf \sqrt{1-z^2} \cos \theta, \sqrt{1-z^2} \sin \theta, z \ri}$.
Taking $J$ to be of order 1, the entire action is of order $S$ in the large $S$ limit.
In this limit we may simplify to a quadratic spin-wave Lagrangian. Let us write this in two ways.

First, it is convenient to write the spin wave Lagrangian as a lattice field theory for a single scalar.
We consider quantum fluctuations on top of a state in which the staggered spin order parameter lies in the $(x, y)$ plane, expanding the action to quadratic order in ${z_j}$ and the fluctuations ${\theta_{j+1}-\theta_j - \pi (1-c_jc_{j+1})/2}$ of the relative angles. It is useful to define the height field (counting field) associated to ${S_z}$:
\begin{align}
h_{j} & = {\sum}_{i=1}^j z_i,
& 
{(\nabla h)_j } &  \equiv h_j - h_{j-1} = z_j. 
\end{align}
We define $h_{0}=0$, while ${h_{L}\times S}$  is the total $S_z$ magnetization and is conserved.
By the result for the ground state spin, there are ground states with $h_L$ ranging from ${-|n_A-n_B|}$ to ${|n_A-n_B|}$ in steps of size~$1/S$. For now let us consider the sector with $S^z_\text{tot}=0$, i.e. $h_L=0$.
This is not a severe restriction, 
since the full many-body spectrum, which falls into SU(2) multiplets, is represented in this sector.

In App.~\ref{app:IBPdetails} we show that integrating out the phase degrees of freedom from Eq.~\ref{eq:initialaction} gives the Lagrangian
 \begin{equation}\label{eq:actionhonly}
 \mathcal{L} = 
\f{S}{2J}
\left(
 \sum_{j=1}^{L-1}  (\partial_t h_{j})^2
+ J^2 \sum_{j,k=0}^L h_j K_{jk} h_k
\right),
 \end{equation}
where we have dropped an additive constant. 
The random symmetric matrix $K$ is defined by

  \be\label{eq:stiffnessenergy}
  \sum_{j,k=0}^L h_j K_{jk} h_k
  =
  \sum_{j=1}^{L-1}
\left[
c_{j+1}(\nabla h)_{j+1} - c_j (\nabla h)_j
\right]^2.
 \ee
We have derived this action at large $S$, neglecting higher-order terms in the expansion around the ordered state: these are RG-irrelevant
for physics near the chosen ground state, although they  are important for the SU(2) symmetry that relates different ground states. 

Second, let us give an alternative formulation directly in the continuum limit, which is useful for heuristic arguments.  Let ${\vec \pi=(\pi^1,\pi^2)}$ be the Goldstone modes in an expansion around a perfectly ordered state, ${\vec N = S(\sqrt{1-\vec\pi^2},\vec\pi)}$.
Then, to leading order at large $S$,
and taking the continuum limit (with unit lattice spacing),  
\be\label{eq:ctmaction}
\mathcal{L} = \f{S}{2} \int \dd x  
\left[
{i} c(x)  {\vec\pi}\times \dot{\vec \pi}
+ J (\nabla \vec \pi)^2
\right].
\ee
Here ${{\vec \pi}\times \dot{\vec \pi}=\epsilon^{ab}\pi^a \dot \pi^b}$,
and $c(x)$ is the continuum analogue of the site-random ${c_j=\pm 1}$.

\tocless\section{Scaling exponents}
The excitations of Eq.~\ref{eq:actionhonly} are quadratic spin waves 
determined by the random kernel $K$.
In general the spin waves are Anderson-localized by disorder, 
but since they are Goldstone modes, the localization length diverges as the energy tends to zero \cite{sheng1990scattering}.
The ground-state correlators of the Goldstone modes are power-laws.

Let us summarize the key exponents before turning to derivations.
The most basic quantity is the dynamical exponent ${z=3/2}$ \cite{PhysRevB.38.4980,gurarie2003bosonic}.
Below we give a simple argument for this value by examining the kernel in Eq.~\ref{eq:stiffnessenergy}.
The dynamical exponent sets the typical energy 
as a function of lengthscale (for the lowest modes, this lengthscale $\ell$ is set by the system size, and for higher modes it is a localization length). In particular the average energy gap for a finite chain scales as 
\ba\label{eq:gap}
\overline{\Delta E} &  \sim J L^{-z}, 
& 
z & = 3/2,
\end{align}
where the bar is the disorder average.
In the following we will compare this with DMRG calculations.
The value ${z=3/2}$ indicates a density of states for spin-wave modes scaling as ${\rho(E)\sim L J^{-2/3}  E^{-1/3}}$~\cite{gurarie2003bosonic}.

By considering the modes we find that
the fluctuation corrections to the ordered moment converge at large $L$ (unlike in the case of the 1D antiferromagnet), with finite-size corrections of typical size $1/\sqrt{L}$,
and that the ground-state correlators for Goldstone fluctuations scale as follows:
\ba\label{eq:staticcorrelators}
\overline{\big\langle\vec N^\perp_i \cdot \vec N^\perp_j \big\rangle} 
& \asymp  |i-j|^{-1/2},
&
\overline{\big\langle \vec S^\perp_i \cdot \vec S^\perp_j \big\rangle} 
& \asymp  |i-j|^{-3/2}.
\end{align}
Here ${\vec N^\perp \equiv \vec \pi}$ denotes the two transverse components of the staggered order parameter.
Similarly  ${\vec S^\perp = (S^y, S^z)}$ are the transverse components of the (\textit{non}-staggered) spin. 
These simple correlators will suffice for our tests of the applicability of spin wave theory. 
We expect that non-equal-time correlators are related to Eq.~\ref{eq:staticcorrelators} by the usual ansatz \cite{sachdev2023quantum}, multiplying by appropriate scaling functions of ${t/|i-j|^z}$.

The exponents above may be obtained by considering the modes of the kernel $K$ in the Lagrangian (\ref{eq:actionhonly}). 
Here we give a schematic picture, while deferring to App.~\ref{app:upperandlowerbounds} a more in-depth discussion (there we present precise upper and lower bounds on the gap and discuss correlators and the order parameter in more detail). The results are  also in agreement with a handwaving renormalization group argument given below.

After imposing the  Dirichlet boundary conditions ${h_L=h_0=0}$ appropriate to the ${S^z_\text{tot}=0}$ sector,
the left-hand side of Eq.~\ref{eq:stiffnessenergy} is written in terms of a truncated matrix ${K'}$ whose  indices run from $1$ to ${L-1}$.
Decomposing $h_j$ into the eigenmodes of $K'$, with eigenvalues ${\{\lambda_\alpha\}_{\alpha=1}^{L-1}}$, gives  a collection of ${L-1}$ harmonic oscillators with excitation energies
\be\label{eq:omegalambda}
E_\alpha = J \sqrt{\lambda_\alpha}.
\ee
We assume that it  suffices to consider the lowest mode (i.e. that higher modes will obey a similar energy scaling, with $L$ replaced by the appropriate localization length).

First, note that the stiffness cost in Eq.~\ref{eq:stiffnessenergy} vanishes for the following height configuration, which we denote~$B_j$ (for ``Brownian''):
\ba\label{eq:Bdefn}
(\nabla B)_j &= c_j,
&
B_0&=0.
\end{align}
Since the ${c_j=\pm 1}$ are uncorrelated random variables, $B_j$ is a random walk.
In a typical disorder realization, $B_L$ is of order $\sqrt{L}$, so $B_j$ does not satisfy the right-hand Dirichlet boundary condition ${B_L=0}$, and therefore does not yield a zero mode of $K'$.
Instead, the lowest eigenstate resembles a version of $B_j$ that has been smoothly deformed in order to satisfy the right-hand boundary condition. 
In place of Eq.~\ref{eq:Bdefn}, we write this lowest (un-normalized) eigenstate $\Psi_j$ in terms of a slowly-varying vector $\mu_j$ (${\mu_1=1}$): 
\ba\label{eq:Psidef}
(\nabla \Psi)_j & = c_j \mu_j, &
\Psi_0 & = 0.
\end{align}
In a typical disorder realization, 
the typical size of $(\nabla\mu)_j$ must be of order~$1/L$
in order to satisfy the boundary condition ${\Psi_L=0}$ (details in App.~\ref{app:upperandlowerbounds}).

The size of the gap follows from this scaling of ${\nabla\mu}$, together with the scaling of the normalization of $\Psi$. The typical order of magnitude of the elements $\Psi_j$ is $\sqrt{L}$, like those of the original random walk $B_j$. 
As a result, the scaling of the lowest eigenvalue~is
\be\label{eq:lowestenergystate}
\lambda_\text{min} =  
\f{(\Psi^TK\Psi)}{|\Psi|^2}
=
\frac{
\sum_j \left[ (\nabla\mu)_{j}\right]^2 
}{
\sum_j \Psi_j^2
}
\sim \f{1}{L^3},
\ee
as stated above. By Eq.~\ref{eq:omegalambda}, this gives the scaling (Eq.~\ref{eq:gap}) for the gap.

We note that the scaling in Eq.~\ref{eq:gap} holds for the average and for typical samples. If we condition on having a sample with an \textit{atypically} small value of ${|n_A-n_B|}$, we obtain a smaller energy gap of order ${J|n_A-n_B|/L^2}$.

A direct extension of the heuristic picture above for the modes also shows that the fluctuation correction to the ordered moment is convergent
--- this is also confirmed numerically below ---
and gives the correlator scalings discussed above (see {App.~\ref{app:sec:orderparam} for further details).

At a heuristic level, the exponents in Eqs.~\ref{eq:gap},~\ref{eq:staticcorrelators} follow more simply from naive dimensional analysis of the Lagrangian (Eq.~\ref{eq:actionhonly}). For example, the second term in the action may be written ${\int \dd t \sum h K h \sim \int \dd t \dd x \, (\nabla N^z)^2}$. 
In order for the action to be dimensionless, and using ${t\sim x^{3/2}}$, we require that $N^z$ has scaling dimension ${\Delta=1/4}$.

We may also consider a renormalization group treatment of the $\vec\pi$ Lagrangian in Eq.~\ref{eq:ctmaction}.
At the handwaving level, this is straightforward.
We wish to imagine a ``Wilsonian'' coarse-graining procedure (by a factor $b$ in lengthscale) under which the action 
${\mathcal{S}=\int \dd t \mathcal{L}}$, which involves quenched randomness,
is statistically invariant.
This coarse-graining effectively averages the random function $c(x)$ over a scale $b$;
in order to ensure that $c(x)$ retains a typical magnitude of order 1, we must also replace ${c\rightarrow b^{-1/2} c}$.
Then to keep the action invariant the other replacements 
must be 
${x\rightarrow b x}$,
${t\rightarrow b^z t}$,
${\pi \rightarrow b^{-\Delta} \pi}$,
with ${z=3/2}$ and ${\Delta=1/4}$, recovering the results above.

Since the Gaussian theory is a simple example of a disordered fixed point with exactly-known exponents,
it would  be interesting to try to make the RG treatment precise. This would require a precise formulation of a ``Wilsonian'' transformation in which some version of Eq.~\ref{eq:ctmaction} was invariant.
Here, we  use the RG  picture only to argue (nonrigorously) that the leading spin wave \textit{interactions} are RG-irrelevant, with RG eigenvalue $y=-1/2$, since they have two additional powers of $\vec \pi$ in $\mathcal{L}$.

\tocless\section{Numerical spin wave diagonalization}

As a check on  the claims above, we make a numerical analysis of the spin wave problem using the standard Holstein Primakoff representation of the Hamiltonian~\cite{auerbach1998interacting}.
Taking the ordering direction to be parallel to $N^x$, we introduce a bosonic mode on each site, whose annihilation operator $a_j$ is related to the ordered component of the spin through $c_j S^x_j = S - a_j^\dag a_j$. 
The boson vacuum is the ``classical'' ground state,  which will be dressed by quantum fluctuations.

To leading order in $1/S$, the Hamiltonian is quadratic in the bosons, and can be diagonalized with a Bogoliubov transformation~\cite{COLPA1978327}. 
However, to regularize the problem, 
it is necessary to  eliminate the zero mode associated with global spin rotations. 
We have considered two setups: 
an open chain with a boundary field applied to the leftmost spin,
and a periodic chain with an ``infinitesimal'' uniform field (in practice of size $10^{-7} J$). 
In the latter case it is necessary to exclude samples with ${n_A=n_B}$, but the fraction of such samples vanishes in the thermodynamic limit.
Details are in App.~\ref{app:sec:spin-wave-theory}.

First we have computed the reduction to the ordered moment due to quadratic spin waves.  
In the limit of large $L$, we find that
\be\label{eq:orderedmomentlargeS}
\overline{\<N^x\>} = S - 0.4564(4) + O(1/S).
\ee
The second term is the large-$L$ limit of ${\overline{\langle a^\dag_j a_j\rangle}}$.
Consistently with the scaling results of the previous section,  we find that these fluctuations are finite as $L\rightarrow\infty$, 
and that the finite size corrections at large $L$ are essentially of order $1/\sqrt{L}$:  see App.~\ref{app:sec:orderparam} and App.~\ref{app:sec:more-spin-wave-data} for a discussion of subleading corrections.

In the inset to Fig.~\ref{fig:spin-wave} we show data for the energy gaps $\Delta_\alpha^{\rm av}$ of the first few (nonzero) spin-wave modes.
The scaling of these gaps is in good agreement with 
the $L^{-3/2}$ discussed above.

Next, the main panel of Fig.~\ref{fig:spin-wave} shows the typical value $G^{\rm typ}_N(r)$ of the correlator $(2S)^{-1}{\langle\vec N^\perp_i \cdot \vec N^\perp_{i+r} \big\rangle}$,
again computed numerically within quadratic spin wave theory. 
The reason for showing the typical, rather than the average, is  to avoid rare-sample effects
discussed in App.~\ref{app:sec:orderparam} and App.~\ref{app:sec:more-spin-wave-data}.  
(The typical is defined in the standard way using the average of the logarithm.)
At large  separation $r$, the scaling of the correlator is close to the $r^{-1/2}$ expected from the previous section.
Finite size effects decay relatively slowly. 
Since our argument for this exponent is heuristic, we cannot rule out logarithmic corrections. 

Finally, power-law correlations hint at a bipartite entanglement entropy scaling like $\log L$. In App.~\ref{app:sec:entanglement} we present numerical results, within linear spin-wave theory, consistent with this scaling.

\begin{figure}
    \centering
    \includegraphics[width=\linewidth]{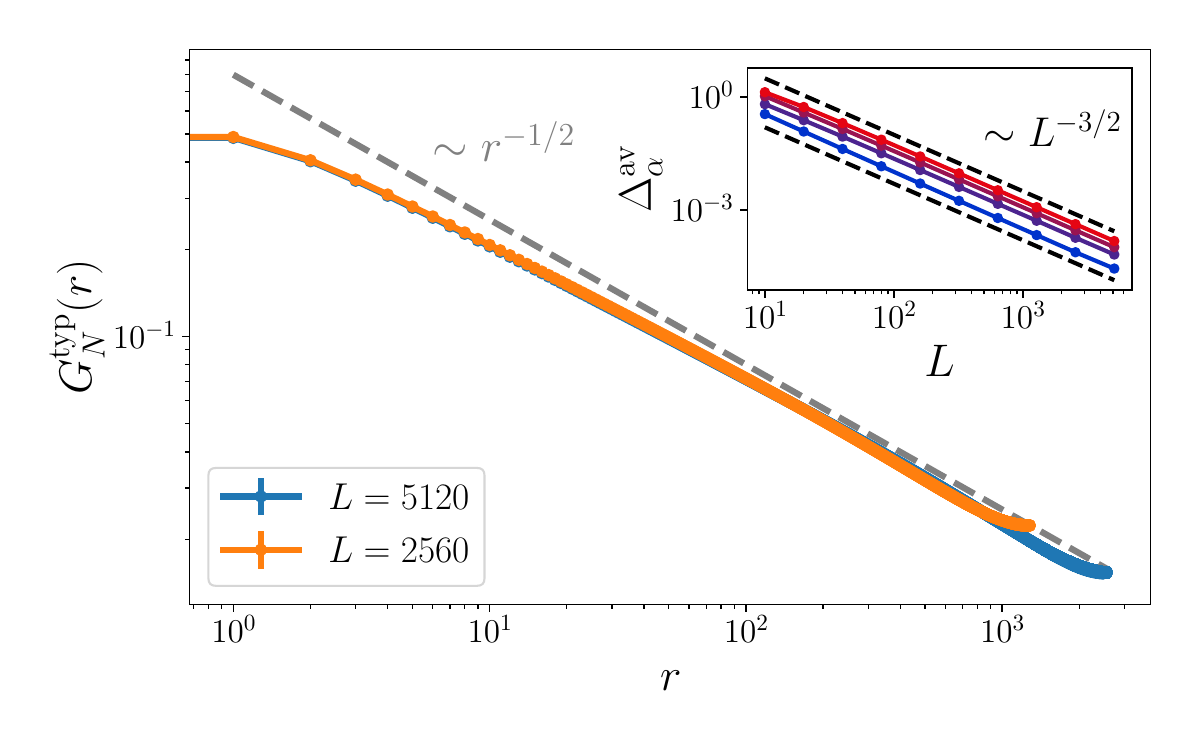}
    \caption{Quadratic spin wave theory: The typical value of the correlator $G_N$ at distance $r$ for the two largest available system sizes. The dashed line indicates the expected power law, ${G_N^{\rm typ}\varpropto r^{-1/2}}$.
    Inset: average energy $E_\alpha$ of the four lowest-lying spin wave modes. (The energy of the zero mode is not shown and is always below $10^{-7}$.) As a guide for the eye, two black dashed lines show power laws ${\varpropto L^{-3/2}}$. The simulations are performed with periodic boundary conditions and $J=1$, excluding samples with ${n_A=n_B}$.}
    \label{fig:spin-wave}
\end{figure}

\tocless\section{DMRG simulations}

We expect spin wave theory to be accurate for nonuniversal quantities only at large $S$,
but to capture \textit{universal} properties more broadly.
We have argued that the interaction terms appearing in the  Lagrangian (Eq.~\ref{eq:actionhonly}) at higher orders in $1/S$ are RG-irrelevant, 
so the above universal behaviour should hold at least for sufficiently large $S$. But the simplest possibility is that this 
universal behavior holds for all $S$. 
Now we investigate this using Density Matrix Renormalization Group (DMRG)~\cite{PhysRevLett.69.2863,PhysRevB.48.10345,SCHOLLWOCK201196}. 

We focus on the cases $S=1/2,\,1,\,3/2$. 
We find good evidence that our theory applies already at $S=1$ and $S=3/2$. 
For $S=1/2$, we give evidence that the ground state is ordered 
(with a small value of the ordered moment)
but we are not able to numerically confirm the universal values of the critical exponents:
we suggest this is due to larger finite size effects at $S=1/2$.

For the DMRG simulations we again restrict to configurations with ${n_A\neq n_B}$.
The ground state then lies in an $\mathrm{SU}(2)$ irrep with nonzero  spin ${S_{\rm tot} = S|n_A-n_B|}$. 
We use Matrix Product States (MPS) invariant under the $\mathrm{U}(1)$ symmetry generated by ${S^x_{\rm tot}=\sum_j S^x_j}$.
We can therefore select a specific ground state within this
multiplet by working in the sector with ${S^x_{\rm tot}=S_{\rm tot}}$ 
(the same ground state would be selected by applying an infinitesimal field in the $S^x$ direction, 
as in the numerical spin-wave computations above). 
This protocol also defines the axis for the order parameter.

\begin{figure}
    \centering
    \includegraphics[width=\linewidth]{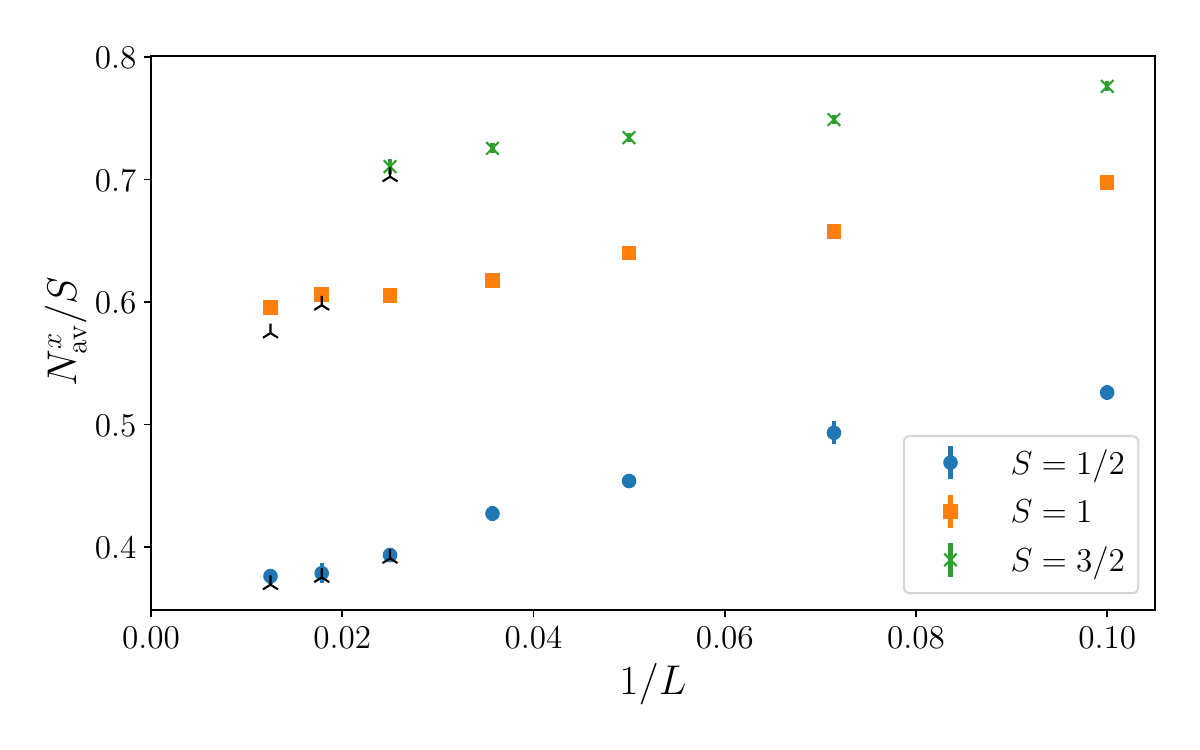}
    \caption{{DMRG data for the order parameter for spin values $S=1/2,~1,~3/2$. The average order parameter $S^{-1} L^{-1}\sum_j \overline{\langle N_j^x\rangle}$ is reported as a function of $1/L$. The data suggests that the order parameter converges to a finite value as $L\to\infty$. 
    DMRG data is converged in bond dimension and number of sweeps; \wye symbols show a conservative estimate of possible systematic error due to rare disorder realizations at large $L$ where DMRG does not achieve a high overlap with the ground state (see App.~\ref{sec:DMRG-convergence}).
    }}
    \label{fig:dmrg-OP}
\end{figure}

Again we start with the average order parameter,
\be
   N^x_{\rm av} = L^{-1} {\sum}_j \overline{   \langle N^x_j \rangle}.
\ee
A minimal requirement for spin-wave theory to be valid is that $N^x_{\rm av}$ remains positive in the thermodynamic limit. 
Data for $S=1/2,\,1,\,3/2$ are shown in Fig.~\ref{fig:dmrg-OP} as a function of the inverse system size $1/L$. In all cases, the data suggests that $N^x_{\rm av}$ converges to a finite value in the thermodynamic limit.
For comparison, spin-wave theory (Eq.~\ref{eq:orderedmomentlargeS}) predicts
$\lim_{S\to\infty}\lim_{L\to\infty}(S-N^x_{\rm av}) \approx 0.46$. Using the largest system sizes available with DMRG we find $S-N^x_{\rm av} \approx 0.31,\,0.40,\,0.43$ for $S=1/2,\,1,\,3/2$ respectively.

Further evidence for long-range order 
comes from correlation functions (data reported in App.~\ref{sec:DMRG-convergence}).
For all $S$, these show a very strong asymmetry between the longitudinal and transverse components of the staggered spin $\vec{N}$, 
and the longitudinal correlator $\overline{\< N^x_j N^x_k\>}$ is consistent with convergence to a nonzero value at large separation. 
We note that a strong-disorder RG study of a chain with $\mathcal{J}_i$ distributed continuously in ${[-J_0,J_0]}$ proposed a non-long-range-ordered state with correlations decaying as ${1/\ln r}$ \cite{PhysRevB.60.12116}. 
On numerical grounds it is difficult to exclude such a slow decay, but we believe that the long range order predicted by spin wave theory 
is a simpler explanation of the data in Fig.~\ref{fig:dmrg-OP}  (see below for further discussion).

\begin{figure}
    \centering
    \includegraphics[width=\linewidth]{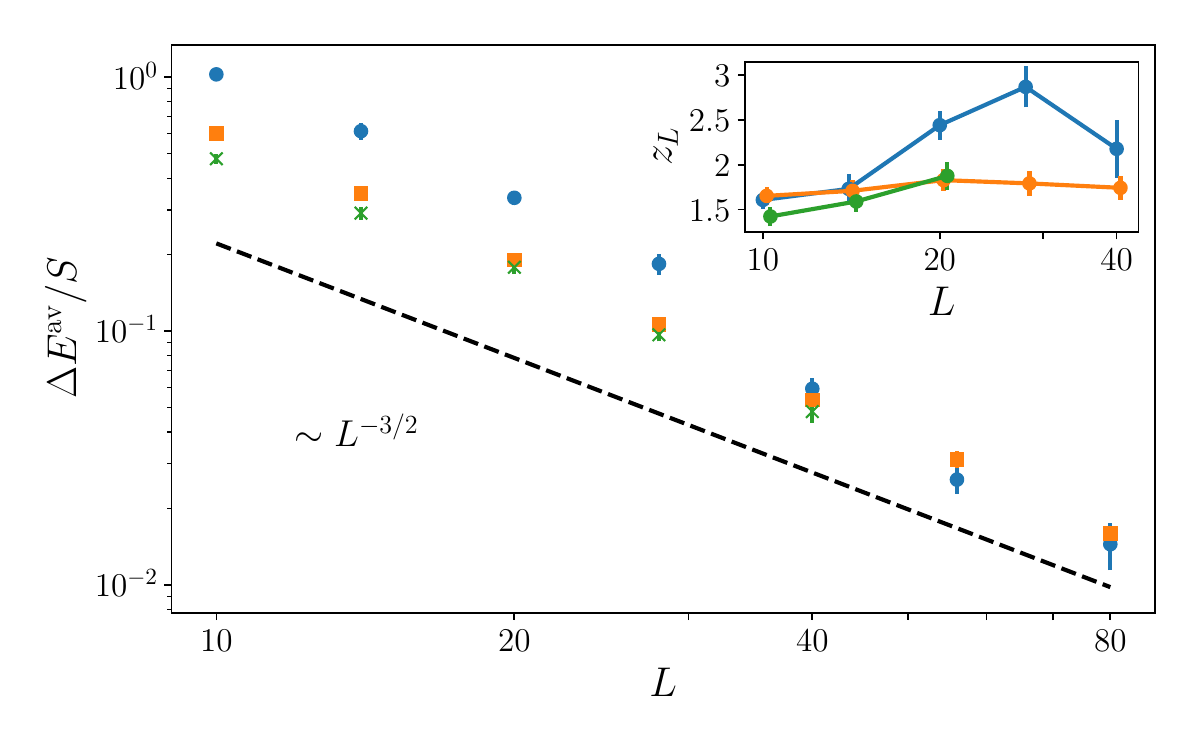}
    \caption{DMRG data for the energy gap for spin values $S=1/2,~1,~3/2$. Main panel: average energy gap $\Delta E / S$ (defined in main text) as a function of $L$. For comparison, dashed line reports the predicted slope for dynamical exponent $z=3/2$. Error bars (smaller than symbol size in many cases) report statistical errors.
    Inset: running estimate of the dynamical critical exponent, defined in the main text.
    }
    \label{fig:spin-one-gap}
\end{figure}

Next we analyze the dynamical scaling exponent $z$. As a representative energy scale we take the  gap ${\Delta E}$ between the ground state (computed as above) 
and the lowest-energy state in the sector with ${S^x_{\rm tot}=S(|n_A-n_B|+1)}$. 
The latter is the lowest-energy state with spin larger than the ground state.
Fig.~\ref{fig:spin-one-gap} shows the dependency of the average gap $\Delta E^{\rm av}$ on the system size.
An initial  impression is that the data for $S=1$ and $S=3/2$ are roughly in agreement with the $L^{-3/2}$ trend line. 
A closer examination shows that it is difficult to obtain a precise numerical estimate of $z$, as a result of finite-size effects on these lengthscales. 
In the inset to Fig.~\ref{fig:spin-one-gap} we show running exponents $z_{L}$,
computed from power-law fits using a given system size $L$ and the next two larger sizes. Note the strong non-monotonic finite size effects for $S=1/2$.
For $S=1$, $z_L$ for the largest $L$ values  is above the expected value $z=3/2$, but we conjecture that there will be convergence to $z=3/2$ for larger sizes.

{
Finally we look at the transverse Goldstone mode correlator. 
Fig.~\ref{fig:GN-spin-one} shows 
the typical value $G_N^{\rm typ}(r)$ for $S=1$,
as a scaling plot. (Raw data for ${S=1/2}$, ${S=1}$, ${S=3/2}$ is reported in App.~\ref{sec:DMRG-convergence}.)
At the available scales,  $G_N^{\rm typ}$  in Fig.~\ref{fig:GN-spin-one} decays faster than the linear spin wave prediction $r^{-1/2}$~(Eq.~\ref{eq:staticcorrelators}). However, it seems likely to us that this is a result of finite-size effects, since, for a given $r$, the  curves become less steep as  $L$ increases. 
(Note that the $r^{-1/2}$ behavior is expected only when ${r\gg 1}$ and ${r/L\ll 1}$ both hold.)
The deviation is even stronger for $S=1/2$
(with decay closer to $1/r$ for $r\sim 10$)
but  there is strong $L$-dependence. Again we think the most likely explanation for the deviation is finite-size effects, i.e. that  at sufficiently large scales the noninteracting spin wave prediction for the exponent would be recovered. 
}

Finally we comment on DMRG convergence. 
In all cases we perform standard checks of convergence  as a function of bond dimension/number of sweeps.
As an independent check that DMRG is not ``stuck'' in a state far from the true ground state, we consider the ratio ${\epsilon = (\langle H^2\rangle-\langle H \rangle^2) / \Delta E^2}$.
For the vast majority of samples
the MPS found by DMRG is close to the ground state (${\epsilon\ll 1}$).  
At the largest sizes, 
a small percentage of rare samples, with a smaller gap, do not achieve small $\epsilon$. 
In App.~\ref{sec:DMRG-convergence} we estimate  the maximum possible error from these samples.
The resulting  cautious lower bound on the true order parameter  was shown as a \wye symbol in Fig.~\ref{fig:dmrg-OP}.
The effect on the gap estimate is negligible compared to the statistical error. 

\begin{figure}[t]
    \centering
    \includegraphics[width=\linewidth]{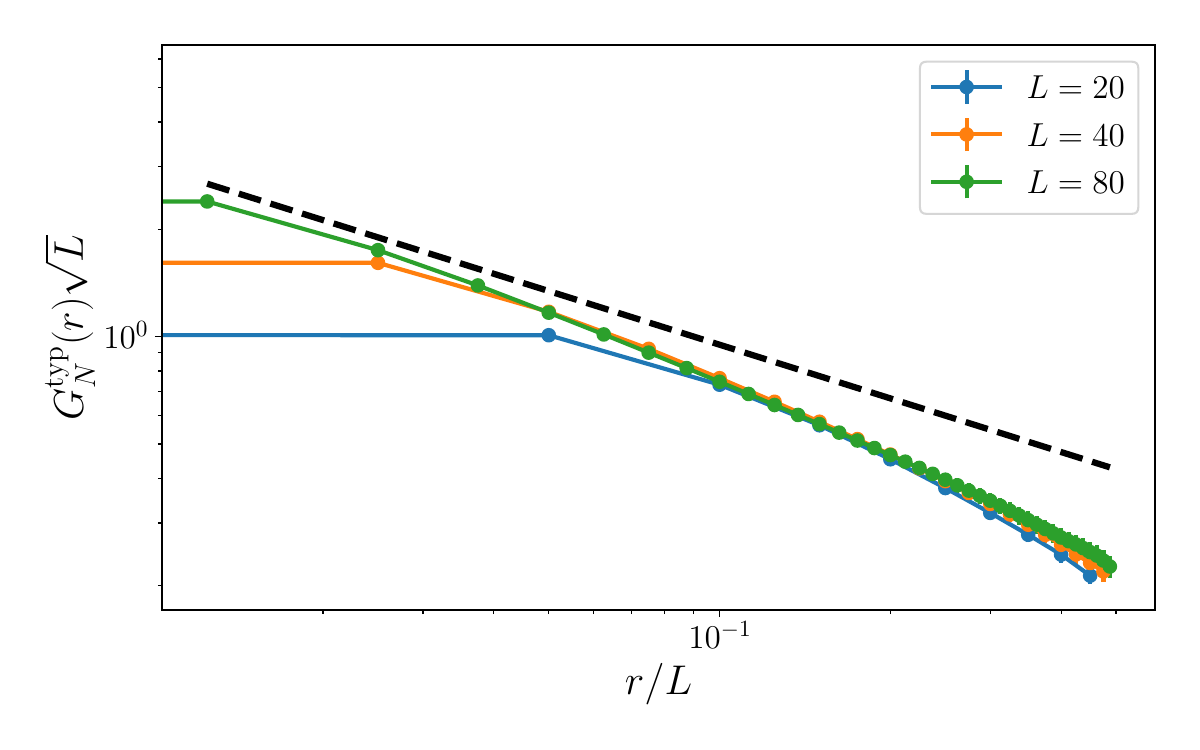}
    \caption{Correlator $G_N^{\rm typ}(r)$ for $S=1$ chains (scaling plot for various $L$). As a guide for the eye, a dashed black line indicates the expected power-law decay $r^{-1/2}$. Data is obtained by averaging over the central $L/2$ sites.}
    \label{fig:GN-spin-one}
\end{figure}

\tocless\section{Outlook and extensions}

We have argued that the isotropic spin-$S$ chains with random ${\pm \mathcal{J}}$ couplings have long-range-ordered ground states, dressed by nontrivial spin wave fluctuations. 
This shows the possibility of nontrivial 
breaking of continuous symmetry \cite{anderson1990broken,peierls1991reflections} in one spatial dimension, 
and is contrary to what was expected previously for this system.

For concreteness 
we assumed $\mathcal{J}_j=\pm \mathcal{J}$ with probability $1/2$, but the scaling theory is more general.
For example, the universal scaling is unchanged if the densities of ferromagnetic and antiferromagnetic bonds are distinct (so long as both are nonzero) 
or if different sites have different spin.
Other applications include purely antiferromagnetic chains with randomness in the local  $S$ value \cite{PhysRevLett.75.4302}
and defect spins in randomly dimerized magnets \cite{kimchi2018valence}.
Long-range correlated
or quasiperiodic signs for the couplings can give different exponents, as is easily seen by generalizing the heuristic RG for Eq.~\ref{eq:ctmaction} to 
an arbitrary scaling 
${\sum_{j=1}^b c_j \sim b^{\xi}}$,
which gives a dynamical exponent ${z=1+\xi}$ (see App.~\ref{app:sec:RG-extension}}).

The universal behavior that we find, in particular the presence of ordering, is different to that predicted by numerical strong-disorder RG~\cite{PhysRevB.52.15930,PhysRevB.60.12116,PhysRevB.55.12578}.
However, it should be noted that our disorder distribution did not allow arbitrarily weak couplings (i.e ${\mathcal{J}\sim 0}$), unlike the standard model studied in the strong-disorder RG literature \cite{PhysRevB.52.15930,PhysRevB.60.12116}. 
Allowing arbitrarily weak bonds is probably innocuous if the small-${|\mathcal{J}|}$ tail of the probability distribution,
${P(|\mathcal{J}|)\dd |\mathcal{J}|\sim |\mathcal{J}|^{a}\dd |\mathcal{J}|}$
is light enough, 
but if the tail is heavy enough then the scalings will change.
 A priori, the ground state properties for sufficiently heavy tails could be closer to strong-disorder RG predictions. 
However, a heuristic RG approach, described in App.~\ref{app:sec:RG-extension}, suggests that long-range order remains stable  for any tail exponent ${a>-1}$, with the dynamical exponent remaining  ${z=3/2}$ for ${a>0}$ and taking the value ${z=(3+a)/(2+2a)}$ for ${a<0}$. 

We also note that finite-size effects are large for the ${S=1/2}$ case, and  simulations for larger sizes would be desirable to better test our hypothesis of long-range order.
On the more formal side, analysis of corrections to scaling both from  (i)~lattice effects in the noninteracting theory 
and (ii)~RG-irrelevant spin-wave interactions would also be interesting, and may shed light on the relatively large finite size effects that we see for the exponents.
It also remains to characterize the full probability distribution of correlation functions (arising from disorder), 
and to formulate precisely the renormalization group for the disordered Lagrangians  (Eqs.~\ref{eq:actionhonly}, \ref{eq:ctmaction}).

In the future, the phenomenological consequences of the theory presented here could be investigated, in the hope of comparing with experiments. Simple thermodynamic consequences include the scaling ${C\sim (T/J)^{2/3}}$ of the specific heat, and the scaling ${\chi\sim S^2/T}$ of the susceptibility (the latter coincides with the scaling within strong-disorder RG \cite{nguyen1996design,PhysRevLett.75.4302}).
The nonlinear sigma model associated with Eq.~\ref{eq:ctmaction} suggests that at finite temperature the correlation length scales as $\xi\sim JS/T$.
Transport and dynamical correlators will be affected by the Anderson localization of the spin-wave modes at the Gaussian level.

Let us comment on variations of the model, some of which could be explored further. 

First, we note that the nontrivial scaling of the random-sign  chain relies on $\mathrm{SO}(3)$ symmetry.
Easy-plane anisotropy gives a term ${n_z^2 \sim (\nabla h)^2}$ in the Lagrangian (Eq.~\ref{eq:actionhonly}): this is strongly relevant according to the scaling dimensions above, and leads to a standard free boson theory in the infrared (IR). This free boson theory describes \textit{quasi}-long-range order of the ``spin glass'' order parameter ${\vec N}$. 
This is therefore a situation where reducing the global symmetry destroys long-range order, rather than enhancing it.
Easy-axis anisotropy presumably immediately gives Ising ``spin glass'' order. 

In the isotropic model, 
we may ask about transitions between the (stable) spin-glass ordered phase and a gapped paramagnetic dimerized phase.
A dimerized phase can be induced by turning on sufficiently strong antiferromagnetic couplings on a subset of bonds --- see App.~\ref{app:sec:transition-to-paramagnet}.
One scenario is that  the initial  ordered state gives way to the paramagnetic state by a sequence of $O(L)$ closely spaced level crossings (at which the spin $S_{\rm tot}$ of the ground state multiplet,  and the spin glass ordering pattern, change).
More generally, it may be interesting to study the effect of strong frustrating further-neighbor couplings on the random-sign chain. 
(Frustrating interactions are RG-irrelevant if they are weak, but may have a nontrivial effect if they are strong.)
The nonstandard scaling properties of the action in Eq.~\ref{eq:ctmaction} may also be a basis for other kinds of unusual criticality when the model is perturbed.

The effective field theory in Eq.~\ref{eq:ctmaction} may be extended to higher dimensions as a description of Heisenberg-Mattis models \cite{sherrington1979long}
 with random-sign, but unfrustrated, couplings.
It could also describe a bipartite antiferromagnet with random site dilution.
However, in higher dimensions we should also include the symmetry-allowed term 
$(\dot{\vec \pi})^2$. This term is irrelevant in 1+1D according to the RG discussed around Eq.~\ref{eq:ctmaction},
but becomes (naively) marginal in 2+1D, while in 3+1D it is more relevant than the term $c(x){\vec \pi}\times \dot{\vec \pi}$. 
It may be possible to explore the competition between the two kinds of time-derivative terms in 2+1D by perturbing in one or the other.
This is different to the approach  in Ref.~\cite{sherrington1979long}, which perturbs around the anisotropic (easy-plane) limit.

Finally, we note that the spin chain has a natural generalization to $\mathrm{SU}(n)$ symmetry 
(which can also be formulated as an anisotropic classical 2D loop model), and it would be interesting to extend the spin wave analysis to general $n$.
Perhaps other symmetries and order parameter manifolds might also lead to interesting ``spin-glass''-ordered states.

\tocless\acknowledgments{We thank 
J. Chalker,
M. Oshikawa,
F. Essler,
T. Senthil,
S. Parameswaran,  
A. Pandey, and D. Kovrizhin
for discussions.
DMRG simulations were performed using the ITensor library~\cite{itensor,itensor-r0.3}. This work was supported by the French Agence Nationale de la Recherche (ANR) under grant ANR-21-CE40-0003 (project CONFICA).}

%

\clearpage

\newpage
\setcounter{equation}{0}
\renewcommand{\theequation}{S\arabic{equation}}
\setcounter{figure}{0}
\renewcommand{\thefigure}{S\arabic{figure}}

\renewcommand{\theequation}{S\arabic{equation}}

\renewcommand{\thefigure}{S\arabic{figure}}

\setcounter{figure}{0}

\setcounter{equation}{0} 

\begin{appendix}

\onecolumngrid

\begin{center}
    \textbf{\large Appendix}
\end{center}
\vspace{2em}

\twocolumngrid

    \tableofcontents

\section{Details of lattice action derivation}
\label{app:IBPdetails}

In this appendix we label the height field $h$ and the relative angular fluctuations $\delta$
by ${(j+\f{1}{2})\in (\mathbb{Z}+\f{1}{2})}$, to emphasize that these quantities live on bonds:
\ba
\delta_{j+1/2} & =\theta_{j+1}- \theta_i - \pi \eta_j,
&
h_{j+1/2}& =\sum_{k=1}^j z_k.
\end{align}
Here ${\eta_j=(1-c_j c_{j+1})/2}$ is ${\eta_j=1}$ for 
for an AF bond and ${\eta_j=0}$ for an F bond.
We will use ${(\nabla h)_j=h_{j+1/2}-h_{j-1/2}}$ for the lattice difference.
Starting with Eq.~{(3)}, we expand in ${z= \nabla h}$ and $\delta$  to quadratic order:
\be\label{eq:appexpandaction}
\mathcal{S}_0= S \sum_j \int \dd t \lf 
- i (1-z_j) \dot \theta_j 
+ \f{J}{2} \lf 
\delta_{j+1/2}^2 
+ 
T_j
\ri
\ri,
\ee
where:
\ba
T_j & = \lf z_{j+1}-(-1)^{\eta_j} \, z_j \ri^2.
\end{align}
The $L$ phase degrees of freedom $\theta_j(t)$ 
may be written in terms of the ${L-1}$ differences $\delta_{j+1/2}(t)$, together with a spatially constant mode $\theta(t)$  [which may be taken equal to $\theta_1(t)$].
It is convenient first to deal with the integral over the spatially constant part, $\theta(t)$. 
Writing ${\mu(t) = \partial_t \theta(t)}$,
this appears in the Berry phase term as
\be
\mathcal{S}_B^\text{zero mode} = - i \int \dd t \mu(t) F(t),
\ee
where $F(t)$ is the number of spin flips with respect  to the all-up state,
\be
F(t) = S \sum_j (1-z_j) = S \lf  1 - h_{L+1/2} \ri.
\ee
If $\mu(t)$ was completely unconstrained, the functional integral over $\mu(t)$  would impose a delta-function constraint $F(t)=0$. 
However, the time integral of $\mu(t)$ is the temporal winding of the phase and is constrained to be  $2\pi$ times an integer:
\be
\int \dd t \mu(t) = \theta(\beta) - \theta(0) \in 2\pi \mathbb{Z},
\ee
Taking this into account, the integral over $\mu(t)$ 
fixes $F(t)$ to be a time-independent integer.\footnote{To see this, we can write the functional integral over $\mu(t)$ as 
\be
\int \mathcal{D} \mu(t) \lf  \sum_{n=-\infty}^\infty e^{i n  \int \dd t \mu(t)} \ri e^{-i \int \dd t \mu(t) F(t)}, 
\ee
where the piece in brackets is a Dirac comb that imposes $\f{1}{2\pi} \int \dd t \mu(t)\in \mathbb{Z}$. Rearranging, this is
\be
\sum_{n=-\infty}^\infty 
\int \mathcal{D} \mu(t) e^{i \int \dd t \mu(t) (F(t) - n) }.  
\ee
Fixing $n$, the $\mu$ integral gives a functional delta  imposing ${F(t)=n}$ for every moment in time. 
Therefore, the  path integral splits into sectors labelled by $n$, and in each sector $F(t)$ is time-independent and equal to $n$.}

Having taken care of the spatial zero mode of $\theta_j(t)$ we can  integrate (or sum) by parts in $t$ and $j$. 
There are no temporal boundary terms because all of the fields are periodic.
(Since we assume the fields are slowly varying in space, the temporal winding number of $\delta_{j+1/2}$ is zero, due to cancellation between $\theta_{j+1}$ and $\theta_j$.)
There are no spatial boundary terms because 
$h_{1/2}=0$ by definition and $\dot h_{L+1/2}=0$ by the constraint above. 

After integrating by parts the piece of the action that depends on the phase fields is then written only through the differences $\delta_{j+1/2}$:
\be
\mathcal{S}_\delta = S \sum_j \int \dd t \lf 
- i  \delta_{j+1/2} (\partial_t h_{j+1/2}) 
+ \f{J}{2} \delta_{j+1/2}^2
\ri.
\ee
Integrating over $\delta_{j+1/2}$, and including the $T_j$ terms from Eq.~\ref{eq:appexpandaction},  gives the action stated in the main text.

\section{Upper and lower bounds on gap}
\label{app:upperandlowerbounds}

We would like to bound the lowest eigenvalue $\lambda_\text{min}$ of the ${(L-1)\times(L-1)}$ matrix $K'$, which satisfies 
\be\label{eq:stiffnessenergyappendix}
\sum_{j,k=1}^{L-1} h_j K_{jk}' h_k
=
\sum_{j=1}^{L-1}
\left[
c_{j+1}(\nabla h)_{j+1} - c_j (\nabla h)_j
\right]^2,
\ee
where on the right-hand-side we take $h_0=h_L=0$.
(We have reverted to the convention in the main text for position indices.)

We revert to the notation in the main text where heights are labeled by ${j\in \mathbb{Z}}$,
writing a given (non-normalized) eigenvector or height configuration $\Psi$ in the form
\ba\label{eq:Psidefapp}
(\nabla \Psi)_j & = c_j \mu_j, &
\Psi_0 & = 0, & 
\Psi_L & = 0.
\end{align}
(An eigenvector of $K'$ has ${L-1}$ components ${\{\Psi_j\}_{j=1}^{L-1}}$, but we complete it to a configuration for $j=0,\ldots,L$ by setting ${\Psi_0=\Psi_L=0}$.)
We will assume that ${n_A\neq n_B}$
(i.e. that ${B_L\neq 0}$),
since the probability that $n_A=n_B$ tends to zero at large $L$.

\subsection{Lower bound}

For a lower bound, we take $\Psi$ to be the lowest eigenstate, $\Psi=\Psi^{\rm min}$, and we show that the requirement that it satisfies the boundary conditions imposes a restriction on the eigenvalue 
\be\label{eq:lambdamintobound}
\lambda_{\rm min} = \f{
\sum_{j=2}^{L} (\nabla\mu)_j^2
}{
\sum_{j=0}^{L} \Psi_j^2.
}
\ee
We will need a lower bound on the numerator and an upper bound on the denominator in (\ref{eq:lambdamintobound}). 
Recall that our convention is $(\nabla\mu)_j = \mu_{j}-\mu_{j-1}$, etc.\footnote{A potentially confusing result of this notation is that $(\nabla\mu)_j$ is associated with the same bond of the spin chain as $\Psi_{j-1}$.}

First let us bound the size of the elements appearing in the denominator. By Eq.~\ref{eq:Psidefapp}, which defines $\mu_k$ (and recalling that $c_k = [\nabla B]_k$) 
we have\footnote{Note that
${ \sum_{k=1}^j \alpha_k (\nabla\beta)_k 
=\alpha_j\beta_j - \alpha_1 \beta_0 - \sum_{k=1}^{j-1} \beta_k (\nabla\alpha)_{k+1}}$. }
\ba\label{eq:psiIBP}
\Psi_j = \sum_{k=1}^{j} \mu_k (\nabla B)_k 
= B_j \mu_j - \sum_{k=1}^{j-1} B_k (\nabla \mu)_{k+1}.
\end{align}
Therefore
\ba
|\Psi_j | 
& \leq 
|B|_{\rm max} 
\lf 
|\mu_j|  + \sum_{k=1}^{j-1} |(\nabla \mu)_{k+1}|
\ri
\\
& \leq 
|B|_{\rm max} 
\lf 
|\mu_j|  + \sum_{k=1}^{L-1} |(\nabla \mu)_{k+1}|
\ri
\\
& = 
|B|_{\rm max} 
\lf 
|\mu_j|  + (L-1)  |\nabla \mu|_*
\ri,
\end{align}
where we have used $|\nabla \mu |_*$ to denote the spatial average of the norm of the gradient.  Also, trivially,
\be
|\mu_j| \leq |\mu_1| + (L-1) |\nabla \mu|_*,
\ee
so 
\ba
|\Psi_j | 
& \leq
|B|_{\rm max} 
\lf 
|\mu_1|  + 2 L  |\nabla \mu|_*
\ri.
\end{align}
This gives us
\be\label{eq:normupperbound}
\sum_j \Psi_j^2 \leq L |B|_\text{max}^2 
\lf 
|\mu_1|  + 2 L  |\nabla \mu|_*
\ri^2.
\ee
Using ${\sum_{j=2}^L(\nabla\mu)_j^2\geq (L-1) |\nabla\mu|_*^2}$, 
Eq.~\ref{eq:normupperbound} gives
\be\label{eq:lambdaintermediatelowerbound}
\lambda_{\rm min} \geq \f{
(L-1) |\nabla\mu|_*^2
}{
L |B|_\text{max}^2 
\lf 
|\mu_1|  + 2 L  |\nabla \mu|_*
\ri^2
}
\ee
Next, we use the boundary condition ${\Psi_L=0}$ to bound the ratio between $|\mu_1|$ and $|\nabla\mu|_*$.
From (\ref{eq:psiIBP}),
\ba
0 = \Psi_L &  = B_L \mu_L - \sum_{k=1}^{L-1} B_k (\nabla\mu)_{k+1} 
\\
& = 
B_L \mu_1 - \sum_{k=1}^{L-1} \lf B_k - B_L \ri (\nabla\mu)_{k+1},
\end{align}
so that
\be\label{eq:mubound}
|\mu_1| \leq L \f{|B-B_L|_{\rm max}}{|B_L|}  |\nabla \mu|_*,
\ee
where ${|B-B_L|_{\rm max}}$ is the largest value of $|B_j-B_L|$.
(Note that, by Eqs.~\ref{eq:normupperbound},~\ref{eq:mubound}, $|\nabla\mu|_*$ is nonzero.)
Combining Eq.~\ref{eq:mubound}  with Eq.~\ref{eq:lambdaintermediatelowerbound},
\ba\label{eq:lambdalowerbound}
\lambda_{\rm min} \geq \f{1}{|B|_{\rm max}^2 L^2}
\f{
1-1/L
}{
\lf 2 + \f{|B-B_L|_{\rm max} }{ |B_L| } \ri^2 
}.
\end{align}
The quantity on the right-hand side depends only on the realization of randomness, which is encoded in the random walk $B_j$. 
When $L$ becomes large (by the usual scale-invariance properties of Brownian motion) the three quantities $|B_L|$,
$|B|_{\rm max}$, 
and $|B-B_L|_{\rm max}$ are all of order $\sqrt{L}$, so that we can write the above in the form
\be
\lambda_{\rm min} \geq \f{C}{L^3} ,
\ee
where the constant $C$ depends on the disorder realization but is of order 1 at large $L$.  This is the desired result.

If we consider atypical samples, in which we enforce an atypically small value of ${|B_L|=|n_A-n_B|}$, we can suppress the gap. This is reflected in Eq.~\ref{eq:lambdalowerbound}, which we can rewrite as
\ba
\lambda_{\rm min} \geq \f{B_L^2}{|B|_{\rm max}^4 L^2}
\times
\f{
1-1/L
}{
\lf 2 \f{|B_L|}{|B|_{\rm max}} + \f{|B-B_L|_{\rm max}}{|B|_{\rm max}}  \ri^2 
}.
\end{align}
If we fix $|n_A-n_B|$ to a value much smaller than $\sqrt{L}$ but choose the disorder realization otherwise uniformly at random, then the right hand side is of order ${|n_A-n_B|^2/L^4}$.

\subsection{Upper bound}

For an upper bound on $\lambda_{\rm min}$, we now take $\Psi$ in Eq.~\ref{eq:Psidefapp} to be a variational state (rather than an eigenstate).
For convenience we will take $\mu_1$ to be close to 1.
We will consider a configuration in which $|(\nabla\mu)_j|$ is of order $1/L$, so that the numerator in the variational bound
\be
\lambda_{\rm min} \leq  \f{
\sum_{j=2}^{L} (\nabla\mu)_j^2
}{
\sum_{j=0}^{L} \Psi_j^2.
}
\ee
is of order $L^{-1}$.
What we need to argue is (A) that in a typical realization it is possible to find such a configuration which obeys the boundary conditions, 
and (B) that the typical value of $|\Psi_j|$ is at least of order $\sqrt{L}$, so that the denominator is of order $L^2$. This then gives us an upper bound of order $L^{-3}$, matching the lower bound in the previous subsection.
We will not attempt to be rigorous.

Since all we are aiming for is the correct power law in the scaling, we choose a variational state of a simple form,
\be
\mu_j = 1 -  \f{j}{L}  u,
\ee
where $u$ is to be determined. From (\ref{eq:Psidefapp}),
\ba
\Psi_L  = \sum_{j=1}^L c_j \mu_j 
=  \sqrt{L} \lf \mathcal{A} - u \mathcal{B} \ri,
\end{align}
where
\ba\label{eq:ABdefns}
\mathcal{A} & =  \f{1}{L^{1/2}} \sum_{k=1}^L c_k ,
&
\mathcal{B} & = \f{1}{L^{3/2}} \sum_{k=1}^L c_k k.
\end{align}
$\mathcal{A}$ and $\mathcal{B}$ depend on the disorder realization,
but both are of order 1 in the large $L$ limit (for example ${\overline{\mathcal{A}}=0}$, ${\overline{\mathcal{B}}=0}$, 
${\overline{\mathcal{A}^2}=1}$, 
${\overline{\mathcal{B}^2}\to1/3}$.)
Typically ${\mathcal{B}\neq 0}$ so we can satisfy the boundary conditions by taking ${u=\mathcal{A}/\mathcal{B}}$.
This establishes (A) above, that, in a typical realization of disorder, we can satisfy the boundary conditions with a $\mu$ whose gradients are of order $1/L$.

Next we must consider $\sum_j \Psi_j^2$. We can write
\be
\Psi_j = j^{1/2} \mathcal{A}_j - u \f{j^{3/2}}{L} \mathcal{B}_j,
\ee
where $\mathcal{A}_j$ and $\mathcal{B}_j$ are defined analogously to (\ref{eq:ABdefns}),
\ba\label{eq:ABdefns2}
\mathcal{A}_j & =  \f{1}{j^{1/2}} \sum_{k=1}^j c_k ,
&
\mathcal{B}_j & = \f{1}{j^{3/2}} \sum_{k=1}^j c_k k.
\end{align}
These are typically of order 1 at large $j$. 
Picking any  $M\leq L$, we can write
\ba
\sqrt{\sum_{j=1}^L \Psi_j^2} & \geq \sqrt{\sum_{j=1}^M \Psi_j^2} 
\\
& \geq   \sqrt{\sum_{j=1}^M j \mathcal{A}_j^2} - \f{u}{L}\sqrt{\sum_{j=1}^M j^3 \mathcal{B}_j^2}.
\end{align}
The first term is of order $M$, while the second is of order $uM^2/L$.
Therefore we can choose an $M$ (of order $L$) such that the right hand side is of order $L$.
This establishes (nonrigorously) point (B), that $\sum_{j=1}^L \Psi_j^2$, for our variational state, is of order $L^2$ in  a typical disorder realization. This gives the desired upper bound on the scaling, $\lambda_{\rm min} < O(L^{-3})$. 
Together with the result of the previous section, the scaling is fixed, 
\be
\lambda_{\rm min} \sim L^{-3},
\ee
and the average  scales the same way. Again, if instead of considering typical disorder configurations we condition on an atypically small value of ${|n_A-n_B|}$ then the scaling of the bound is reduced to ${|n_A-n_B|^2/L^4}$ (because $u$, and therefore $\nabla\mu$, is reduced).

{

\section{Scaling of order parameter and correlators}
\label{app:sec:orderparam}

We use the scaling of the mode frequencies discussed in the main text to argue that the order parameter ${\overline{\<N_j^x\>}=\overline{c_j\<S^x_j\>}}$ remains finite (at zero temperature) in the thermodynamic limit.
We follow the usual procedure of assuming an ordered ground state, and checking that the fluctuation corrections to the moment remain finite.

A given sample has a multiplet of $2{S_{\rm tot} + 1}$ ground states.
First consider the ground state polarized along the ${\pm S^x}$~axis, so that the local reduction in the ordered moment is 
\ba
S - c_j \< S^x_j\> & = S - S c_j \< n^x_j\> 
\\
& =  S \< (n^z_j)^2 \> + O(1/S)
\\
& =  S \< (\nabla h)_j^2 \> + O(1/S)
\end{align}
where we  have expanded $n^x={c_j(1-[n^y]^2-[n^z]^2)^{1/2}}$ in the small fluctuations, and used the symmetry between the two fluctuation directions.
The right hand side is of order 1, since $h$ is of order $1/\sqrt{S}$
(Eq.~(5)).

In the Holstein Primakoff approach the polarized ground states above are the natural ones. 
In discussing the $h$ field in the main text we considered a different ground state, with $S^z=0$. 
This is the uniform superposition of all the polarized ground states whose moment lies in the $(S^x,S^y)$ plane.
However, the two types of ground state should give the same result for the reduction in the ordered moment in the limit ${L\to\infty}$.
For simplicity we now consider the $S^z_{\rm tot}=0$ ground state, for which $h$ obeys Dirichlet boundary conditions.

The expectation value may be expanded in the normalized eigenmodes $\hat \psi^\alpha_j$ of the matrix $K'$ defined in the main text. Omitting the site index, 
\ba\label{eq:fluctuations}
\< (\nabla h)^2\>
& = 
\f{J}{S} \sum_\alpha \int_{-\infty}^{\infty} \f{\dd \omega}{2\pi}\f{(\nabla \hat \psi^\alpha)^2}{\omega^2 + \Omega_\alpha^2}
= \f{J}{2 S} \sum_\alpha \f{(\nabla \hat\psi^\alpha)^2}{\Omega_\alpha}
\end{align}
where $\Omega_\alpha$ is the mode frequency.
Recall that a mode $\alpha$ of frequency $\Omega$ has a typical localization length $\ell_\alpha$ of order ${\ell(\Omega)\equiv (J/\Omega)^{2/3}}$.

We make the following crude simplifications of Eq.~\ref{eq:fluctuations}. We assume that the numerator $(\nabla\hat\psi^\alpha)^2$ is negligible unless the localization center of the mode $\alpha$ is within $\sim \ell(\Omega_\alpha)$ of the chosen site $j$.
If the localization center is within this range, then we estimate ${(\nabla\hat\psi^\alpha)^2\sim \ell(\Omega_\alpha)^{-2}}$,
by a simple extension of the picture for the  lowest mode described in the main text.\footnote{That is, we assume that, prior to normalization, and on scales smaller than the localization length, the mode $(\psi^\alpha)_j$ locally resembles a random walk with steps of order 1 size, and that the squared norm is determined by random walk scaling: ${|\psi^\alpha|^2 = \sum_j (\psi^\alpha)_j^2}$ is of order ${\ell(\Omega)^2}$. This normalization factor appears as a denominator in ${(\nabla \hat \psi^\alpha)^2}$.} With these approximations,
\ba\label{eq:fluctuations2}
\< (\nabla h)_j^2\>
& \sim \f{J}{S} \sum_\alpha \f{\mathds{1}\lf x_\alpha \in [-\ell(\Omega_\alpha),\ell(\Omega_\alpha)] \ri}{\Omega_\alpha \ell(\Omega_\alpha)^2},
\end{align}
where the numerator is an indicator function ensuring the localization center $x_\alpha$ is in the required spatial region.
Letting $\rho(\Omega,x)$ be the density of states for modes with frequency $\Omega$ and localization center at $x$, this is 
\ba\label{eq:fluctuations3}
\< (\nabla h)_j^2\>
& \sim \f{J}{S} 
\int \dd \Omega \dd x  \rho(\Omega,x)
\f{\mathds{1}\lf x \in [-\ell(\Omega),\ell(\Omega)] \ri}{\Omega \ell(\Omega)^2}.
\end{align}
A naive scaling analysis, transforming the above sum to an integral using the expected exponents for $\rho$ and $\ell$, indicates that,  for a typical sample,
\be\label{eq:fluctuationstypicalscaling}
\< (\nabla h)_j^2\>
\simeq 
C_j
- \f{D_j}{S\sqrt{L}}
\ee
where the disorder-dependent constants $C_j$ and $D_j$ are of order 1 size.

The scaling above, giving a finite result for ${\langle (\nabla h)_j^2\rangle}$ when $L\to \infty$,
is the main conclusion of this appendix. However, we note that care must be taken in comparing the average and the typical. 
If we consider the   \textit{average} value of the LHS of (\ref{eq:fluctuationstypicalscaling}), we obtain an anomalously large contribution from rare samples with small ${|n_A-n_B|}$.
This is due to the contribution to (\ref{eq:fluctuations2}) from the lowest mode, which scales as
\be
\< (\nabla h)_j^2\>_{\text{lowest mode}}
\sim \f{J}{S L^2 \Omega_{\rm min}}.
\ee
We argued above that samples with small ${|n_A-n_B|}$ have ${\Omega_{\rm min}\sim J |n_A-n_B|/L^2}$, giving
\be\label{eq:fluctuationsraresamples}
\< (\nabla h)_j^2\>_{\text{lowest mode}}
\sim \f{1}{S |n_A-n_B|}.
\ee
Typical samples have $|n_A-n_B|$ of order $\sqrt{L}$.
However if we average the right hand side of (\ref{eq:fluctuationsraresamples}) over all samples we will get a divergence from rare samples with ${n_A=n_B}$.
If (as in some of our numerics) we exclude the samples with ${n_A=n_B}$ then the average may be checked to be of order $(\log L)/(S\sqrt{L})$, so that the finite size correction to the average is larger, by $\log L$, than Eq.~\ref{eq:fluctuationstypicalscaling}.
The detailed form of the subleading finite-size corrections to the ordered moment will depend on the choice of ground state or boundary conditions. However we find that the form ${a + b L^{-1/2}\log L- c  L^{-1/2}}$ is also consistent with numerical results for the average reduction in the ordered moment in the polarized ground state, for a sample with periodic boundary conditions (see Fig.~\ref{fig:number-size-PBC} in App.~\ref{app:sec:more-spin-wave-data}).

The scaling of correlation functions may be estimated in a similar manner to the discussion of Eq.~\ref{eq:fluctuations}, 
retaining only the contribution of modes with a localization length comparable with or larger than the separation between the points.
(In order to obtain the correlator of the \textit{non}-staggered transverse fluctuations, it is easiest to start with $\overline{\< (h_j - h_k)^2 \>}$, whose scaling is easily fixed using the random-walk picture for the modes, and then take lattice derivatives to obtain $\langle S^z_j S^z_k\rangle$.)
This gives the scalings quoted in the main text. The rare sample effect described in the previous paragraph also affects the \textit{average} correlators in a finite sample when the separation of the points is of order $L$. 
The resulting logarithmic enhancements are seen in the data in App.~\ref{app:sec:more-spin-wave-data}.
To avoid this rare sample effect, in the main text we showed the ``typical'' values of the correlators, defined by disorder-averaging the logarithm of the quantum expectation value. For example,
\be
\big\langle
\vec{N}^\perp_i  
\cdot 
\vec{N}^\perp_j 
\big\rangle_{\rm typ}  = 
\exp
\left[\, 
\overline{
\ln  \big\langle \vec{N}^\perp_i  
\cdot 
\vec{N}^\perp_j  \big\rangle
} \,
\right].
\ee

\section{Extensions of the heuristic RG}
\label{app:sec:RG-extension}

In the main text we described a ``Wilsonian'' RG for the random quadratic spin wave Lagrangian 
\be\label{eq:ctmactionapp}
\mathcal{L} = \f{S}{2} \int \dd x  
\left[
{i} c(x)  {\vec\pi}\times \dot{\vec \pi}
+ J (\nabla \vec \pi)^2
\right].
\ee
Our aim in this Appendix is to generalize this to the cases mentioned in the Outlook section: (\textit{i}) models with correlations in the $c_j$, so that $\sum_jc_j
\sim b^\xi$ with $\xi\neq 1/2$;
(\textit{ii}) models with a nontrivial distribution of bond magnitudes $J_j = |J_j|$. 
However, our considerations below will be heuristic, so should be taken only as motivations for conjectures.
We expect that more rigorous results could be obtained by generalizing the approach in Sec.~\ref{app:upperandlowerbounds}.

We will handle the two modifications (\textit{i}) and (\textit{ii}) together. For (\textit{ii}), we consider the case where $J_j\in [0,J]$ with a distribution
\be
P(J_j ) \dd J_j = \f{1+a}{ J^{1+a}}
(J_j)^a \, \dd J_j. 
\ee
For convenience we will use a lattice formulation of the RG here, starting with the action $(x\in \mathbb{Z}$)
\be
\mathcal{S} = \f{S}{2} \int \dd t \sum_{j}
\left[
{i} c_j  {\vec\pi}_j \times \dot{\vec \pi}_j
+ J_j (\nabla \vec \pi)_j^2
\right],
\ee
where $\nabla$ is a lattice derivative.

We consider coarse-graining by a factor ${b\gg 1}$, aiming to obtain an effective action for the modes on lengthscales $\gtrsim b$.
We introduce a coarse-grained field $\widetilde{\vec \pi}_{\tilde j}$, where the coarse-grained coordinate $\tilde j$ labels blocks of sites.
We make the following crude approximations in the coarse-graining:

For the time-derivative term ($\sum_{j\in \tilde j}$ indicates the sum over the $b$ sites inside a given cell)
\ba\notag
\sum_j c_j \vec \pi_j \times \dot{\vec \pi}_j 
 =
\sum_{\tilde j} \sum_{j\in \tilde j} c_j \vec \pi_j \times \dot{\vec \pi}_j
  \rightarrow & 
\sum_{\tilde j} \bigg(\sum_{j\in \tilde j} c_j\bigg)  \widetilde {\vec \pi}_{\tilde j} \times \dot{\widetilde {\vec \pi}}_{\tilde j} \\
= & \,
b^{\xi} \sum_{\tilde j} 
\widetilde c_{\tilde j} 
\widetilde {\vec \pi}_{\tilde j} \times \dot{\widetilde {\vec \pi}}_{\tilde j} 
\notag
\end{align}
We have defined ${\widetilde c_{\tilde j} ={b^{-\xi} \sum_{j\in \tilde j} c_j}}$ so that the typical magnitude of $\widetilde c_{\tilde j}$ remains of order 1 as the coarse-graining factor $b$ is increased.

For the space-derivative term, we consider two cases. First consider the case where $J$ is non-random (i.e. $a\rightarrow\infty$). Then we approximate
\ba\label{eq:wilsongradient}
J \sum_j  (\nabla \vec \pi)_j^2 
\rightarrow 
J  b^{-1} \sum_{\tilde j} 
 (\widetilde \nabla \widetilde {\vec \pi})^2_{\tilde j},
\end{align}
where $\widetilde\nabla$ is the lattice derivative for the coarse-grained lattice.
This reproduces the standard Wilsonian scaling. 
Here, it amounts to assuming that a mode on lengthscales $\gtrsim b$ can be treated as smooth on much shorter lengthscales. We will assume that the scaling with $b$ in Eq.~\ref{eq:wilsongradient}
continues to hold if the exponent $a$ is large enough.
By comparing with the alternative possibility below, 
we expect that the above scaling holds if $a>0$.

For broader distributions, we expect that the gradient energy no longer scales as in Eq.~\ref{eq:wilsongradient}. 
In the limit of a very broad $J_j$ distribution,
the cheapest way to vary the field by an amount  ${\delta {\vec \pi} = \widetilde{\vec \pi}_{\tilde j + 1} - \widetilde{\vec \pi}_{\tilde j}}$
is to concentrate the gradient on the weakest (microscopic) bond in the region of size $b$. 
We assume this gives the correct scaling of the gradient energy for sufficiently negative $a$. 
The weakest bond in the region has a typical magnitude of order $\sim b^{-1/(1+a)}$, 
so we define the quantity ${\widetilde J_j = b^{1/(1+a)} J_\text{min}}$ which is of order $b^0$.
The above picture gives the following scaling for the gradient action cost: 
\ba\label{eq:nonwilsongradient}
 \sum_j J_j (\nabla \vec \pi)_j^2 
\rightarrow 
b^{-1/(1+a)} \sum_{\tilde j} \widetilde J_{\tilde j}
 (\widetilde \nabla \widetilde {\vec \pi})^2_{\tilde j}.
\end{align}
(The probability distribution for $\widetilde J_{\tilde j}$ is again a power law near the origin with tail exponent $a$.)
Comparing with Eq.~\ref{eq:wilsongradient}, we see that concentrating the gradient on the weakest bond is more favorable than distributing it uniformly when $a<0$.
Therefore in general we assume the coarse-grained form
\ba\label{eq:eithergradient}
 \sum_j J_j (\nabla \vec \pi)_j^2 
&\rightarrow 
b^{-m(a)} \sum_{\tilde j} \widetilde J_{\tilde j}
 (\widetilde \nabla \widetilde {\vec \pi})^2_{\tilde j},\\
 m(a) & = \operatorname{max} \lf 1, \f{1}{1+a}\ri ,
\end{align}
where we assume that $\widetilde J_{\tilde j}$ remains of order 1 after coarse-graining.
Finally, we introduce a rescaled time coordinate, $\tilde t = t/b^z$, giving
\be
\mathcal{S} \rightarrow \f{S}{2} \int \dd \tilde t \sum_{\tilde j}
\left[
b^\xi {i} \widetilde c_{\tilde j}  \widetilde {\vec\pi}_{\tilde j} \times \dot{\widetilde {\vec \pi}}_{\tilde j}
+ b^{z-m(a)} \widetilde J_{\tilde j} (\widetilde \nabla \widetilde {\vec \pi})_{\tilde j}^2
\right].
\ee
We see that in order to have asymptotic scale invariance at large $b$ we must choose
\be
z = m(a) + \xi = \operatorname{max}\lf 1, \f{1}{1+a} \ri+ \xi
\ee
and we must rescale the field ($\widetilde {\vec \pi}\rightarrow b^{-x}\widetilde {\vec \pi}$) with the scaling dimension $x = \f{\xi}{2}$. This gives the results stated in the Outlook.

\section{Transition to a paramagnetic phase }
\label{app:sec:transition-to-paramagnet}

First, as a toy model, let's consider a simple nearest-neighbor Hamiltonian that interpolates between (1) a random-sign chain and (2) a dimerized chain in the spin-Peierls phase:
\be
\mathcal{J}_j =  \mathcal{J}_j^{\rm random} -  \f{1 - (-1)^j}{2} 
\mathcal{J}^{\rm AF}.
\ee
We take random signs ${\operatorname{sign}(\mathcal{J}_j^{\rm random})=\pm 1}$. 
For reasons that will be clear below,  we also take the \textit{magnitudes} of $\mathcal{J}_j^{\rm random}$ in this toy model to be random: for concreteness, we take 
${|\mathcal{J}_j^{\rm random}|\in[\mathcal{J},2\mathcal{J}]}$ for some fixed $\mathcal{J}$, with a uniform distribution in this range. 
We will increase $\mathcal{J}^{\rm AF}$, starting from $\mathcal{J}^{\rm AF}=0$.

When $\mathcal{J}^{\rm AF}=0$ 
we have a random-sign chain which is assumed to be long-range ordered, as argued in this paper, with
\be\label{eq:orderingpattern}
\operatorname{sign}\< {\vec S}_j \>\cdot \<{\vec S_1}\> = c_j c_1,
\ee
where the sublattice labels are 
\be\label{eq:cdefnappendix}
c_j =
\prod_{k<j}
\lf \operatorname{sign} \mathcal{J}_k \ri.
\ee
In the opposite limit, where $\mathcal{J}^{\rm AF}$
is much larger than the maximum value of $|\mathcal{J}_j^{\rm random}|$, the ground state is a trivial product of spin singlets on the even-numbered bonds.

When $\mathcal{J}^{\rm AF}$ is increased from zero, it will flip the sign (one by one) of those odd-$j$ bonds  that are initially ferromagnetic. 
Each time this occurs, 
the nature of the ordered state changes, as, by (\ref{eq:cdefnappendix}), we must redefine the sublattice labels to the right of the flipped bond.
The spin of the ground state $S_{\rm tot}$ also changes (typically by an amount of order $\sqrt L$, and of either sign).
Eventually, after $O(L)$ events, when ${\mathcal{J}^{\rm AF}= 2 \mathcal{J} - O(1/L)}$, every odd bond is antiferromagnetic.
The ground state is then a spin singlet.\footnote{Each odd-$j$ bond connects two sites with opposite values of $c$, so that (for even $L$) ${S_{\rm tot} = S| \sum_{j=1}^L c_j | = 0}$.} We expect this chain to be in the same phase as the trivial  singlet product state. 

During this process (as $\mathcal{J}_{\rm AF} \rightarrow 2\mathcal{J}$), the magnitude of the spatially-averaged  ``spin-glass'' order parameter will gradually decrease to zero,
because (loosely speaking) we are producing larger and larger dimerized antiferromagnetic regions which presumably resemble, locally, a gapped paramagnetic state.

The above model is non-generic because at each level-crossing point (when a coupling is zero) the chain is disconnected.
To obtain a more generic model, 
we can also (for example) add  weak second-neighbor couplings (with any sign structure) between sites $j$ and $j+2$, for $j$ even. The chain is then never disconnected.
An alternative model is one in which we augment the nearest-neighbor random-sign chain with antiferromagnetic bonds of strength $\mathcal{J}^{\rm AF}$ between sites $(j,j+2)$
for $j=1 \operatorname{mod} 4$ and $j=2 \operatorname{mod} 4$, 
giving a product of singlets on these bonds when $\mathcal{J}^{\rm AF}$ is large.
In these more generic models,
a possible scenario is that there is a similar phenomenology to that described above, with a sequence of level crossings leading eventually to the singlet-product state. However, this requires further examination.

In the toy model it was necessary to include randomness in $|\mathcal{J}_j^{\rm random}|$ in order to 
avoid all the level crossings occurring at once.
However, we do not expect this pathology to be an issue in 
in the more generic model.
Even in the  absence of randomness in $|\mathcal{J}_j^{\rm random}|$, the  $\mathcal{J}^{\rm AF}$ value of a given  level crossing event will take a nontrivial random value that depends on the random environment of the bond.

\section{Details of the linear-spin-wave computation}
\label{app:sec:spin-wave-theory}

The ordered state that we are expanding around  has ${S^x_j = S c_j}$. As in the case of the antiferromagnet, 
it is convenient to make change of basis for the spins in the $B$ sublattice, so that this classical ground state becomes ${\tilde {\vec S}_j=S(1,0,0)}$. The spin operator $\tilde{\vec{S}}_j$ is defined as
\ba
(\tilde{S}^x_j,\tilde{S}^y_j,\tilde{S}^z_j) =
\begin{cases}
(S^x,S^y,S^z) & \text{if } j\in A,\\
(-S^x,S^y,-S^z)& \text{if } j\in B.
\end{cases}
\end{align}
($A$ and $B$ sites are those with ${c_j=1}$ and ${c_j=-1}$ respectively.)
We then employ the Holstein-Primakov transformation to describe quantum fluctuations on top of the classical ground state. At lowest-order this yields ${\tilde{S}^x_j=S-a_j^\dag a_j}$, 
${\tilde{S}^+_j { \equiv \tilde S^y+i\tilde S^z} \simeq \sqrt{2S} a_j}$, giving a quadratic Hamiltonian 
\be
\mathcal{H}_q = J \sum_{j=1}^{L-1}
\left[
a_j^\dag a_j^{\phantom{\dag}}  +a_{j+1}^\dag a_{j+1}^{\phantom{\dag}}
+ 
\left\{
\begin{array}{ccc}
- ( a_j^\dag a_{j+1}^{\phantom{\dag}} +   \mathrm{h.c.})  \\
(a_j a_{j+1} + \mathrm{h.c.} )
\end{array}\right\}
\right],
\ee
where the first line in the braces applies for a ferromagnetic bond and the second line for an antiferromagnetic bond. In the following we will set $J=1$.

This Hamiltonian may be written in the form
\ba
\label{eq:H-spin-wave}
\mathcal{H}_q &= \vec{\alpha}^\dag  D \vec{\alpha},
&
\vec{\alpha} &=
\begin{pmatrix}
\vec{a}_A\\
\vec{a}_B^\dag
\end{pmatrix}
\end{align}
where $\vec a_A$ and $\vec a_B$ are column vectors containining the annihilation operators on the A and B sublattices respectively.
The Hamiltonian $\mathcal{H}_q$ can be completely diagonalized by finding a $L\times L$ matrix $T$, s.t.: (i)
\ba
\vec{\beta} &= T \vec{\alpha} = 
\begin{pmatrix}
\vec{b}_C\\
\vec{b}_D^\dag
\end{pmatrix}
\end{align}
defines bosonic operators $b$, i.e. satisfying bosonic commutation relations, and (ii) in terms of $\vec{\beta}$ the Hamiltonian takes the form
\be
\mathcal{H}_q = \vec{\beta}^\dag  \Delta \vec{\beta}, \quad \Delta\text{ diagonal}.
\ee
Here $\vec{b}_C$ and $\vec{b}_D$ are again column vectors containing $n_A$ and $n_B$ operators respectively. The two conditions above translates into
\ba
\label{eq:bogoliubov-conditions}
\left( T^{-1} \right)^\dag D T^{-1} &= \Delta,
&
T J_0 T^\dag &= J_0,
\end{align}
where
\be
J_0 =
\begin{pmatrix}
\mathds{1}_{n_A}& 0\\
0 & -\mathds{1}_{n_B}
\end{pmatrix},
\ee
and $\mathds{1}_{l}$ is the $l\times l$ identity matrix.

To find $T$ numerically, we use the procedure outlined in Secs.~2 and~3 of Ref.~\onlinecite{COLPA1978327}, which we summarize below.
We assume that $D$ is positive definite --- we will comment further on this assumption in the following --- and perform a Cholesky decomposition of $D$ to find a matrix $K$ s.t. $D = K^T K$,%
\footnote{By construction $K$ is upper diagonal, but this is irrelevant for the purpose of the algorithm we are describing.}
with $A^T$ denoting the transpose of $A$.
\footnote{We are using that in our problem $D$ is real. In the case of a complex $D$, all the steps remain the same upon replacing $\bullet^T$ with $\bullet^\dag$.}
We can then form the symmetric matrix $K J_0 K^\dag$ and proceed to diagonalize it, i.e. find a diagonal matrix $\Lambda$ and an orthogonal one $O$ s.t.
\be
K J_0 K^{ T} = O \Lambda O^T.
\ee
At this point it is easy to check that the matrices
\ba
T &= \Delta^{-1/2} O^T K,
&
\Delta = J_0 \Lambda
\end{align}
satisfy the conditions~\eqref{eq:bogoliubov-conditions}.

From $T$ and $\Delta$ one can compute, e.g., one-point and two-point functions on top of the ground state $\ket{GS}$ using the fact that it satisfies $\vec{b}_{C}\ket{GS} = \vec{0}$ and $\vec{b}_{D}\ket{GS} = \vec{0}$. For example, we can study deviations from the ordered state as 
\ba
n_j &:= \left\langle \tilde{S}^x_j - S \right\rangle = \left\langle a_j^\dag a_j \right\rangle\nonumber\\
&= \left( T^{-1} B_0 \left( T^{-1} \right)^{ T} -B_0 \right)_{j,j},
\end{align}
with
\be
B_0 = \begin{pmatrix}
0& 0\\
0 & \mathds{1}_{n_B}
\end{pmatrix}.
\ee
[Here with a slight abuse of notation we denote with $\alpha_j$ the component of $\vec{\alpha}$ associated to site $j$, rather than the $j$-th component of the vector as ordered in (\ref{eq:H-spin-wave}). A similar caveat applies to the indices labelling matrix elements.]
We will further be interested in transverse fluctuations of the spin
\ba
G_{S}(j,l) &:= \f{\left\langle S^y_j S^y_l + S^z_j S^z_l \right\rangle}{2S} = \f{\left\langle \alpha_j^\dag \alpha_l + \text{h.c.} \right\rangle}{2} \nonumber\\
&= \left( T^{-1} B_0 \left( T^{-1} \right)^{ T} \right)_{j,l} \text{ for }j\neq l.
\end{align}
From $G_{S}$ we can further compute two-point functions of $N$:
\be
G_{N}(j,l) = c_j c_l G_{S}(j,l).
\ee

Note that the matrix $D$ associated to $\mathcal{H}_q$ in Eq.~\eqref{eq:H-spin-wave} is not going to be positive definite since there will always be a zero mode associated to global $\mathrm{SU}(2)$ rotations of the spins. We resolve this issue by adding  a field that stabilizes the classical ground state with ${S^x_j= c_j S}$. 
In doing this, we must ensure that the effect of the added field is negligible for the physical quantities of interest. We work with two setups.

The first setup is that of a chain with open boundary conditions and a finite field $h$ is applied to the left-most spin, viz.
\be
\mathcal{H}_q \mapsto \mathcal{H}_q + h a_1^\dag a_1.
\ee
This setting physically transparent: it is clear that a boundary field cannot affect thermodynamic properties in the $L\to\infty$ limit, except for defining a direction for the order parameter. We use this setup when studying~${\sum_j n_j}$.

A second setup is more convenient in all other cases, e.g. when studying the gap or correlation functions. Here we take a chain with periodic boundary conditions and apply a uniform field in the $S^x$ direction, i.e.
\be
\mathcal{H}_q \mapsto \mathcal{H}_q + h c_j a_j^\dag a_j.
\ee
with $\operatorname{sign} h = \operatorname{sign} (n_A - n_B)$ to ensure that $D$ is positive.
For finite $h$ the field will affect the physics of the spin chain. To make sure that the effect of $h$ is negligible we must then check that the energy of the lowest-lying mode $\Delta_0\sim h$ is much smaller than the energy of the first excited state $\Delta_1$. In this way the only effect of the field is to gap out the zero-energy mode without producing additional excitations. In practice we found that ${h=10^{-7}}$ yields good results for $L\leq 5120$.

In this second setting, for the zero-energy mode to acquire a positive energy it is necessary to have $n_A\neq n_B$. 
Therefore, for this setup,
we restrict ourselves to disorder realizations satisfying this criterion. This is also consistent with the way we perform the DMRG simulations, in which we fix the direction of the order parameter 
by selecting the global $S^x$ charge sector with $n_A-n_B$; in doing so we discard configurations with $n_A=n_B$, which would not allow us to fix the orientation of the order parameter in this simple way.

\begin{figure}
\centering
\includegraphics[width=\linewidth]{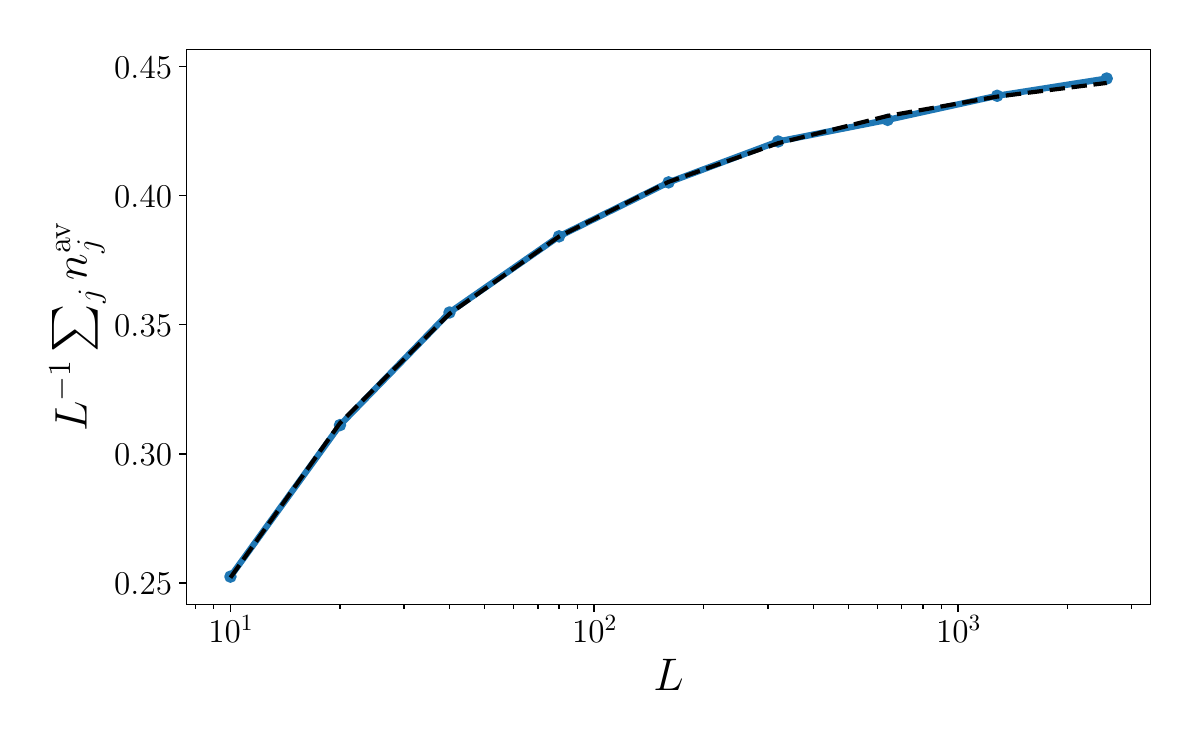}
\caption{The reduction
$n_j=\langle a_j^\dag a_j \rangle$  in the ordered moment, averaged over sites and disorder realizations in the boundary field setup (i.e. a boundary field $h=J=1$ is applied to the leftmost site of a chain with open boundary conditions). The dashed black line reports the best fit of the form $n_0 - p L^{-1/2}$, with $n_0=0.4564(4)$, $p=+0.646(3)$,  and a minimum $\chi^2$ per degree of freedom (d.o.f.) of $\approx0.75$.
$n_0$ is finite, as required for linear-spin-wave theory to be self-consistent.}
\label{fig:number-size}
\end{figure}

\begin{figure}
\centering
\includegraphics[width=\linewidth]{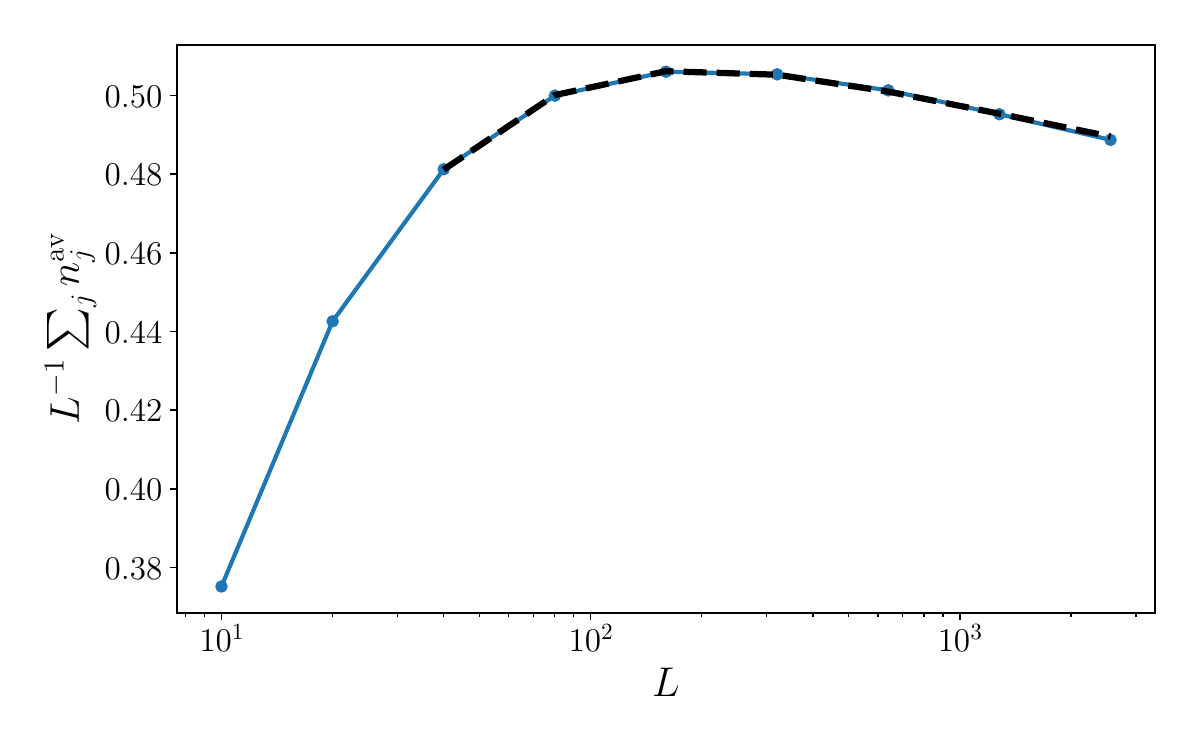}
\caption{ Ordered moment reduction $n_j=\langle a_j^\dag a_j \rangle$ averaged over sites and disorder realizations in the uniform field setup with periodic boundary conditions. The dashed black line reports the best fit of the form $n_0' + (q \log L - p') L^{-1/2}$ with $n_0'=0.4604(7)$, $q=0.323(5)$, and $p'=1.060(1)$, although, with a minimum $\chi^2$ per degree of freedom (d.o.f.) of $\approx1.7$.
}
\label{fig:number-size-PBC}
\end{figure}

\begin{figure}
\centering
\includegraphics[width=0.9\linewidth]{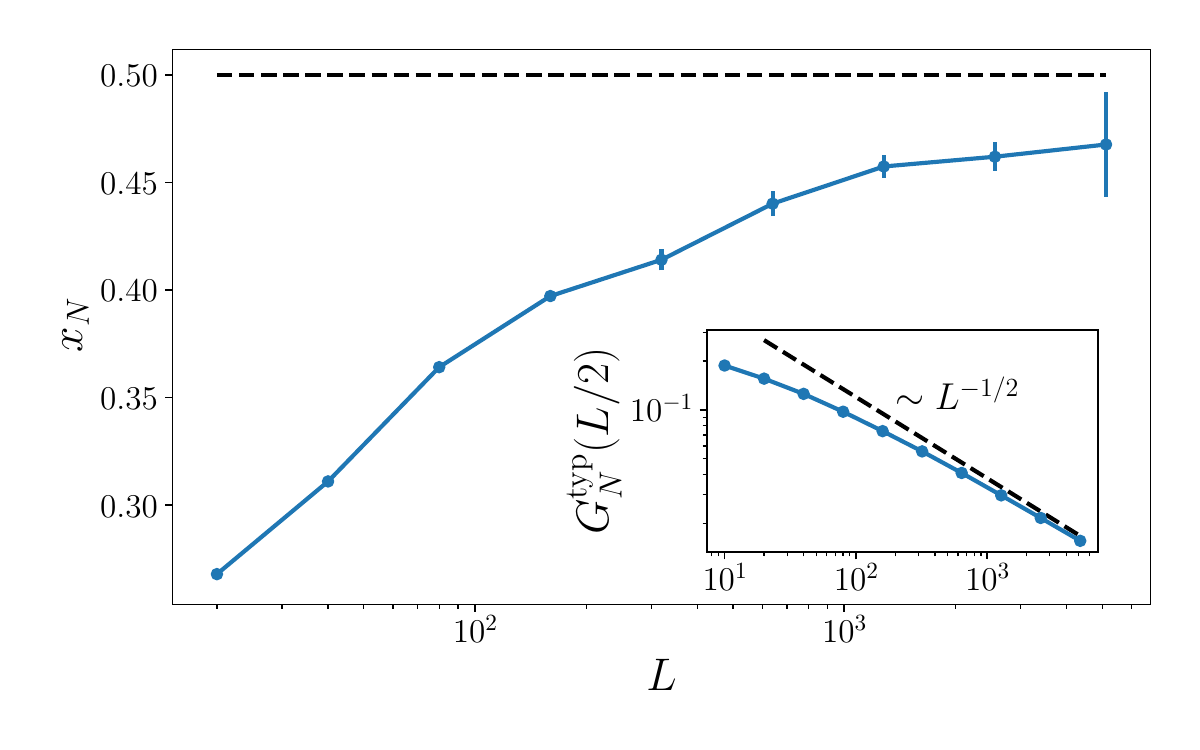}    \includegraphics[width=0.9\linewidth]{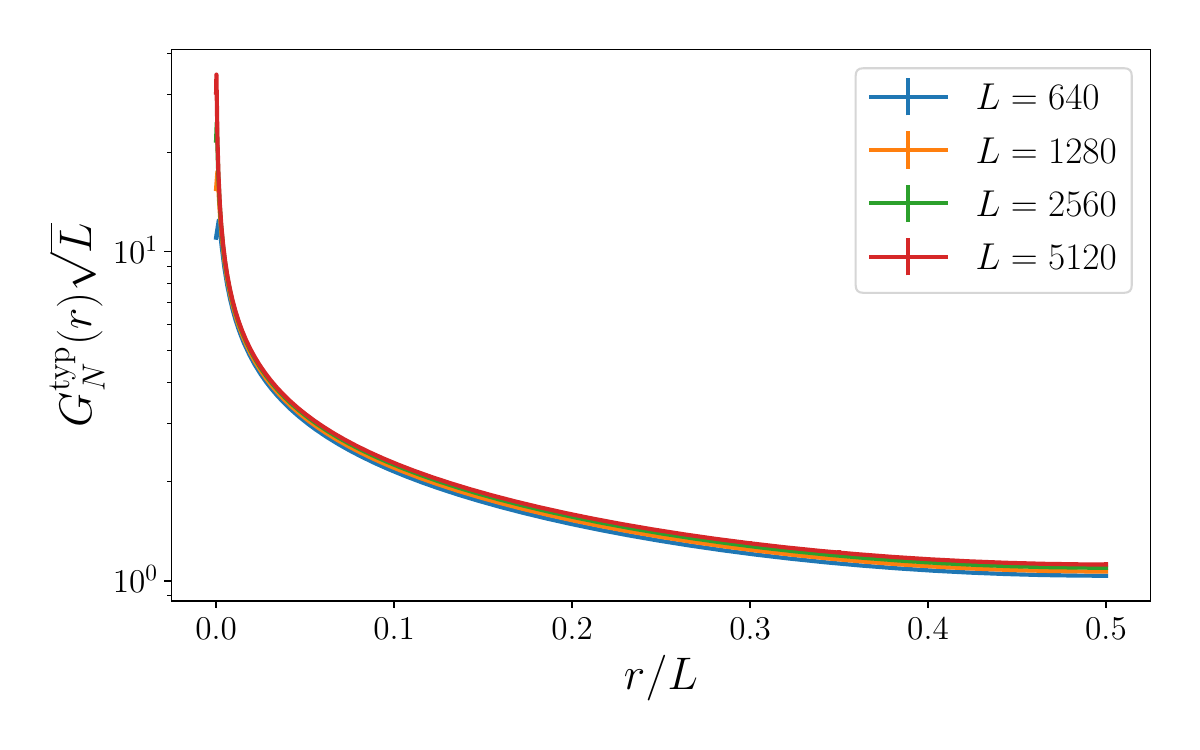}
\caption{Top: the effective finite-size effective exponent $x_N$ estimated from finite differences of $\log G(r=L/2)$ between $L$ and $L/2$. A dashed line shows the value of $x_N=0.5$. An inset shows the data from which the finite differences are computed. Bottom: rescaling the finite-size correlator as $L^{1/2} G_N^{\rm typ}(r)$ yields a reasonable scaling collapse. For graphical convenience errorbars are shown only for for ten values of $r/L$.}
\label{fig:Gtyp-scaling}
\end{figure}

\section{Further data from spin wave computations}
\label{app:sec:more-spin-wave-data}

In this section we report additional numerical data obtained from the numerical linear-spin-wave computation.

First of all, we can ask whether linear-spin-wave theory will correctly describe the physics of the spin chain for some large enough spin $S$.
For the low-order expansion in Holstein-Primakoff bosons to be quantitatively accurate, a necessary condition is that ${n_j= \langle a_j^\dag a_j \rangle \ll S}$ for all sites $j$.
This condition can always be satisfied for large enough $S$ as long as $n_j$ does not diverge in the thermodynamic limit.

In the following we check that there is no such divergence in our model. More precisely, we will show that the spatial and disorder average of $n_j$ remains finite in the thermodynamic limit. This is a nontrivial check, as, for example, it would correctly detect that an antiferromagnetic spin chain cannot be described by linear-spin-wave theory.
To test this condition we work in the boundary-field setup described in the previous section.
In Fig.~\ref{fig:number-size} we report $L^{-1} \sum_j n_j^{\rm av}$. Fitting the data with a function of the form
\be
L^{-1} \sum_j n_j^{\rm av} = n_0 - p L^{-1/2},
\ee
we obtain $n_0=0.4564(4)$, with a minimum $\chi^2$ per degree of freedom (d.o.f.) of $\approx0.75$.
The number in parenthesis denotes the statistical uncertainty on the last digit.
(The change in the $n_0$ estimate due to dropping the smallest two system sizes from the fit was also in the last digit.)
As anticipated $n_0$ is finite and therefore we can expect linear spin-wave theory to be accurate for~${S\gg 1}$.

For completeness, in Fig.~\ref{fig:number-size-PBC} we perform the same analysis in the uniform field setup. As anticipated in App.~\ref{app:sec:orderparam}, here, we expect the finite-size corrections to take a different form 
\be
L^{-1} \sum_j n_j^{\rm av} = n_0' + (q \log L - p') L^{-1/2},
\ee
where, in principle, $n_0'=n_0$.
We can check this is qualitatively compatible with the numerical data for $L\gtrsim 40$. The two estimates $n_0$ and $n_0'$ differ by $\sim 1\%$, which is larger than the statistical error in the fitting parameters, presumably as a result of subleading finite size corrections not included in the fits.

Next we report additional numerical data for the transverse correlation functions. All of the following simulations are performed with a uniform field and periodic boundary conditions, as described at the end of the previous section. In this setting, correlation functions depend only on the distance $r$ between the two sites and on the system size $L$. 
In Fig.~\ref{fig:Gtyp-scaling} we look at the typical value of $G_N(r)$ (we will discuss the mean value below).

In  order to extract the scaling dimension $\Delta$ it is convenient first to examine a correlator that depends on a single lengthscale, so first we examine $G_N^{\rm typ}(r=L/2)$ as a function of system size $L$ (Fig.~\ref{fig:Gtyp-scaling}, Left). The inset shows the raw data.  The main panel shows the effective size-dependent exponent
\be
x_N(\tilde{L}) := \f{\log\left( G^{\rm typ}_N(L/2)\big\lvert_{L=\tilde{L}/2} - G^{\rm typ}_N(L/2)\big\lvert_{L=\tilde{L}} \right)}{\log 2},
\ee
defined so that, if ${G_N^{\rm typ}(r=L/2)\sim L^{-x_N}}$ asymptotically for large $L$, then ${x_N(L)\to x_N}$ as $L\to\infty$. As we show in the left panel of Fig.~\ref{fig:Gtyp-scaling}, $x_N(L)$ converges very slowly. For the system size we have access to $x_N(L)$ still displays large finite-size effects. In spite of this, the data suggests an asymptotic value close to the predicted ${x_N=2\Delta = 1/2}$. 

In the right panel of Fig.~\ref{fig:Gtyp-scaling} we show that rescaling the correlation function as $\sqrt{L}G_N^{\rm typ}$ yields a reasonable scaling collapse across all values of the scaling variable~$r/L$.

Finally, we focus on the average correlators $G^{\rm av}_N$ and $G^{\rm av}_S$ at ${r=L/2}$. As discussed in App.~\ref{app:sec:orderparam}, average correlators scale differently to typical ones once the lengthscale $r$ is comparable with the system size $L$. 
This is due to rare configurations with $|n_A - n_B|\ll \sqrt{L}$ which enhance the average correlators by a factor $\log L$. We show this is the case in Fig.~\ref{fig:Gav-scaling-v2} where we find that the numerical spin-wave computation is consistent with a scaling of the form
\ba
G_N^{\rm av}(r=L/2) &\sim \frac{\log L}{\sqrt{L}}\,,  & 
G_S^{\rm av}(r=L/2) &\sim \frac{\log L}{L^{3/2}}.
\end{align}

\begin{figure}
\centering
\includegraphics[width=0.9\linewidth]{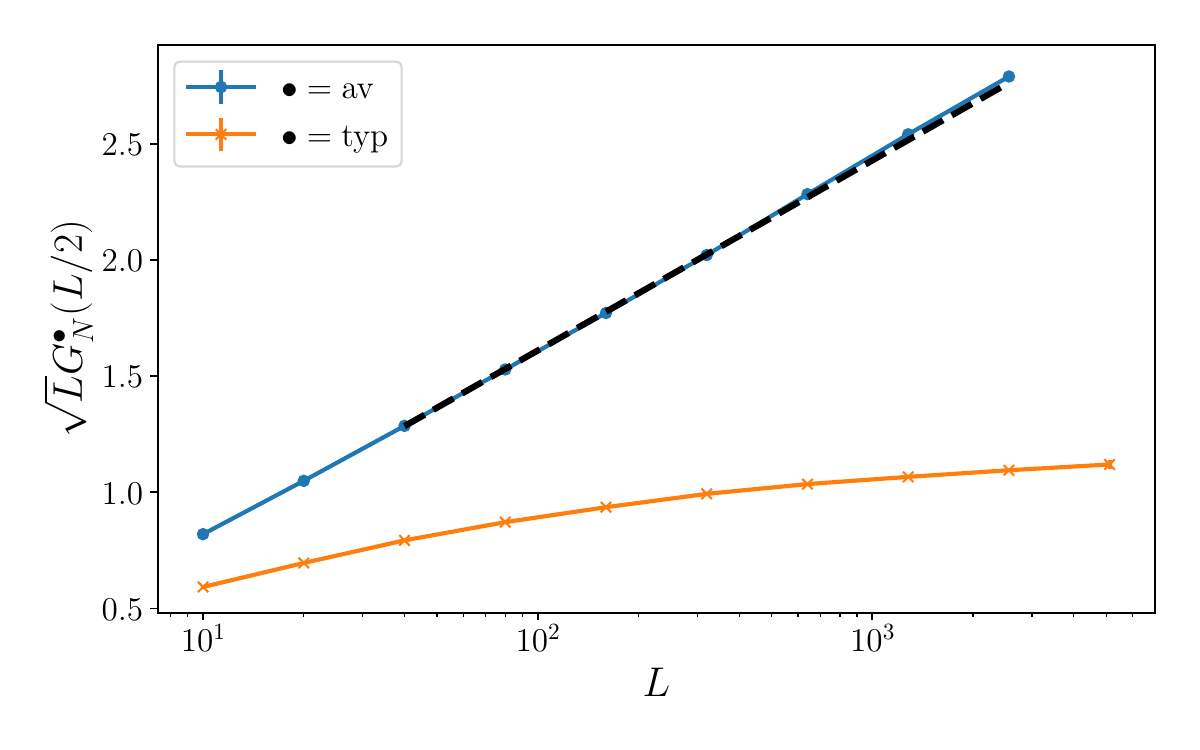}
\includegraphics[width=0.9\linewidth]{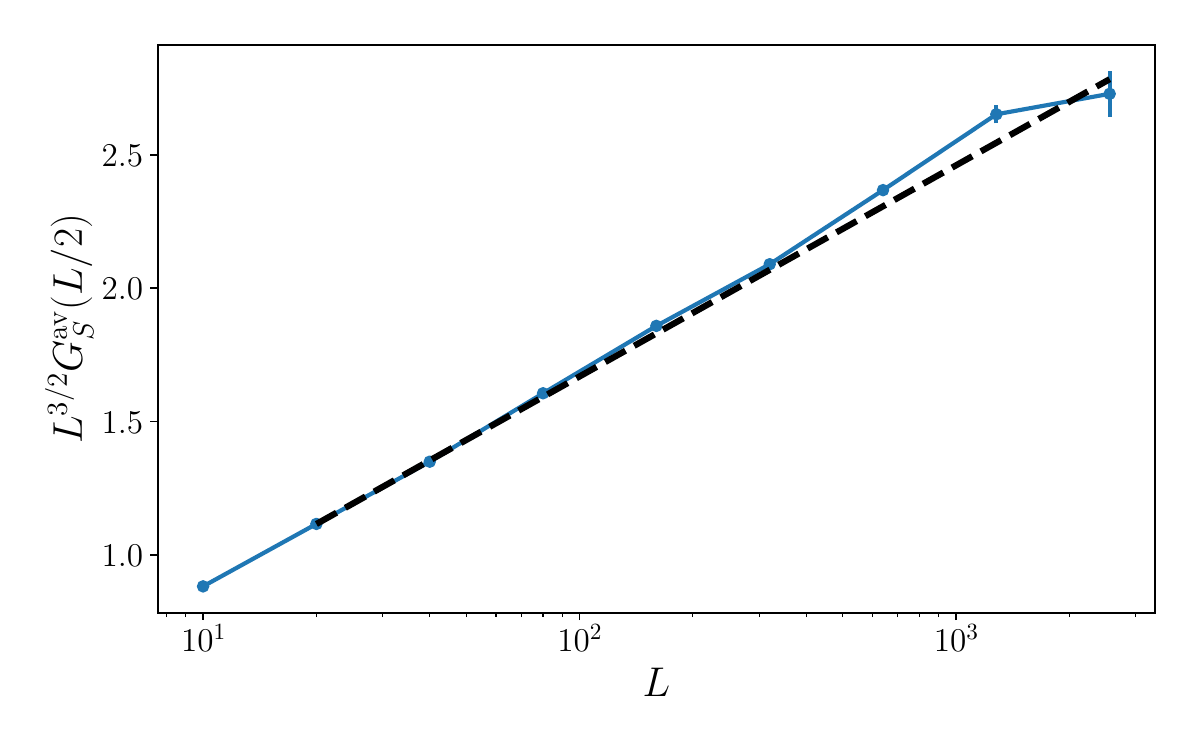}
\caption{$\sqrt{L}G_N^{\rm av}(L/2)$ and $\sqrt{L}G_N^{\rm typ}(L/2)$ (top) and $L^{3/2} G_S^{\rm av}(L/2)$ (bottom) as a function of $L$. As a guide for the eye, black dashed lines report a fit of the form $a+b \log L$.}
\label{fig:Gav-scaling-v2}
\end{figure}

\begin{figure}[h]
\centering
\includegraphics[width=\linewidth]{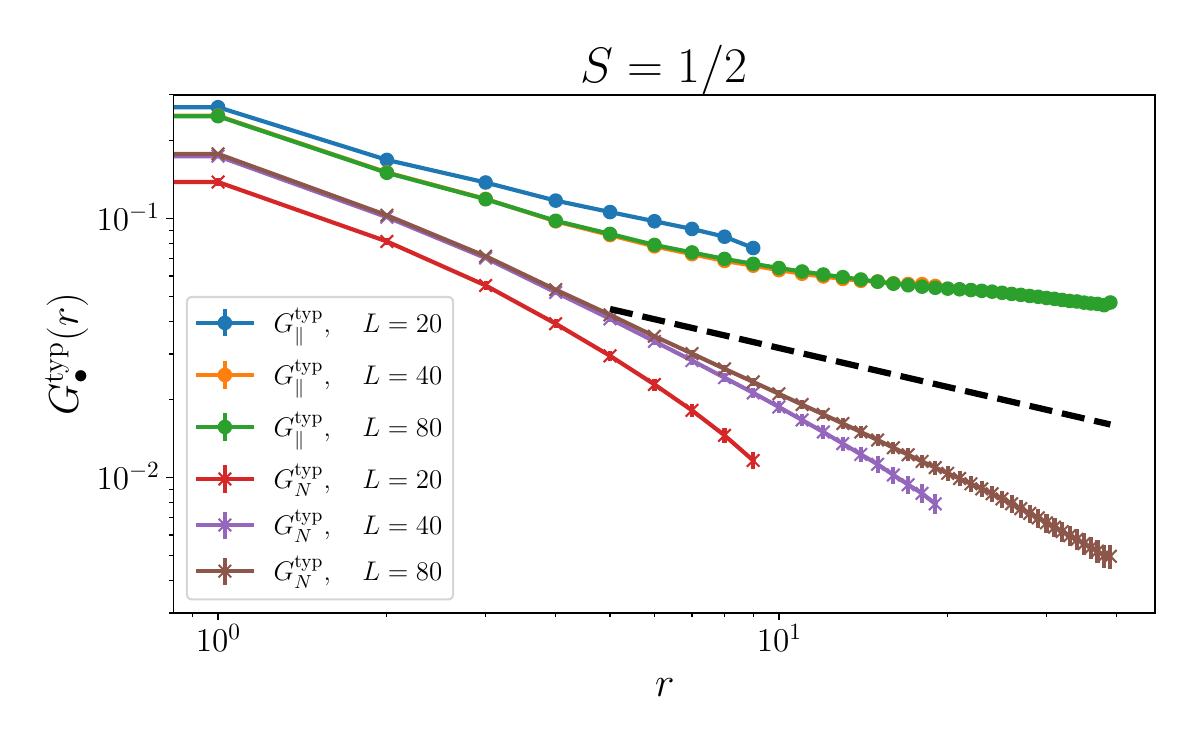}
\includegraphics[width=\linewidth]{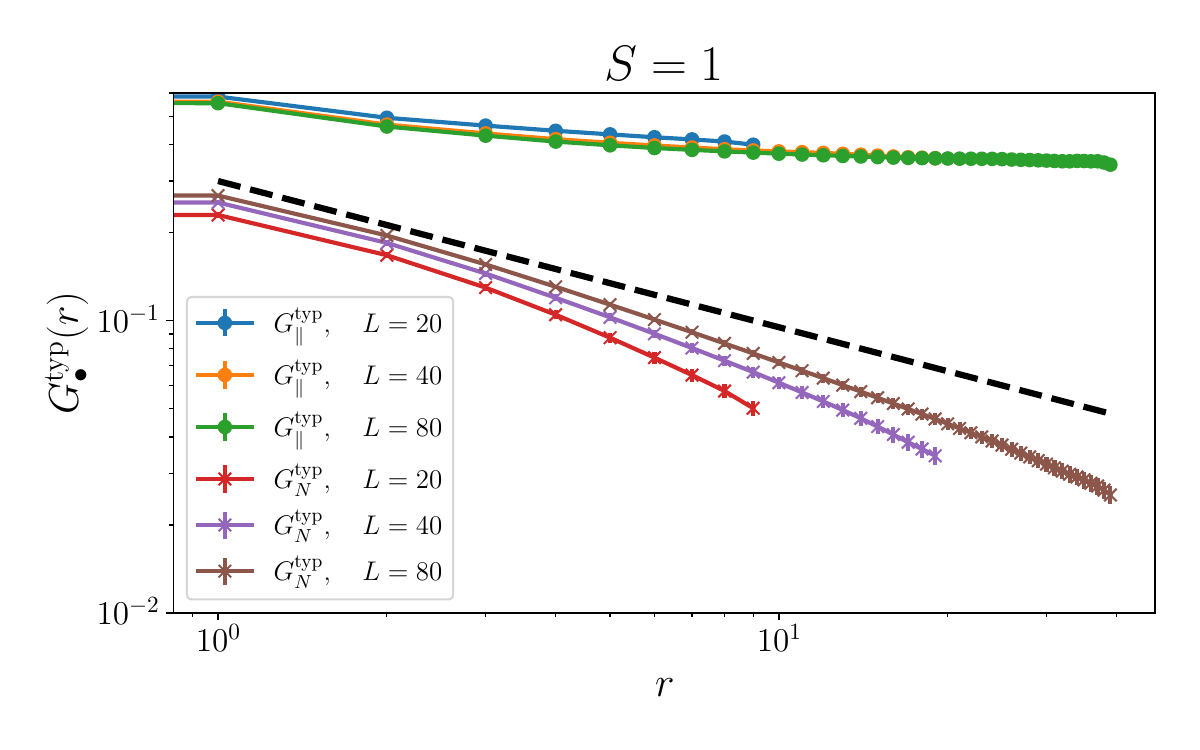}
\includegraphics[width=\linewidth]{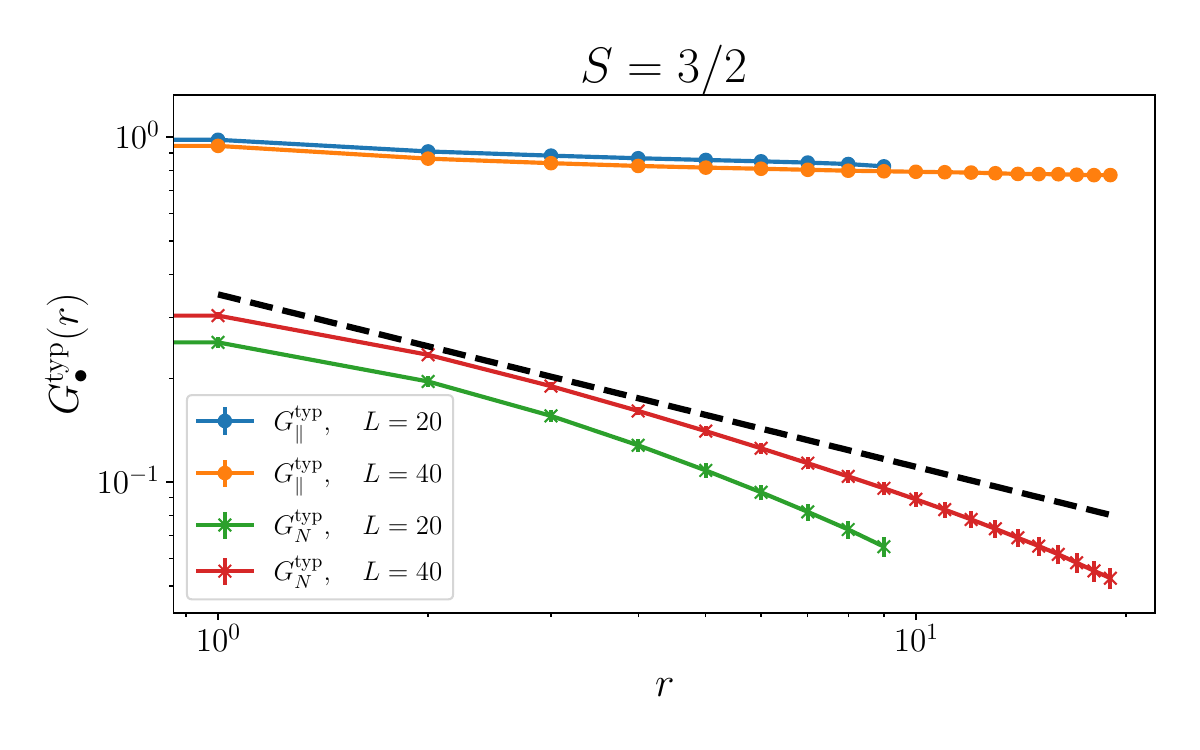}
\caption{Comparison of longitudinal with transverse correlation functions, for various values of the spin (from top to bottom: $S=1/2$, $S=1$, $S=3/2$). Each plot shows both $G_{\parallel}^{\rm typ}(r)$ and $G_N^{\rm typ}(r)$. 
Black dashed lines give a comparison with the slope for $r^{-1/2}$ predicted for $G_N^{\rm typ}$ by linear spin wave theory.}
\label{fig:comparecorrelators}
\end{figure}

\begin{figure}
\centering
\includegraphics[width=\linewidth]{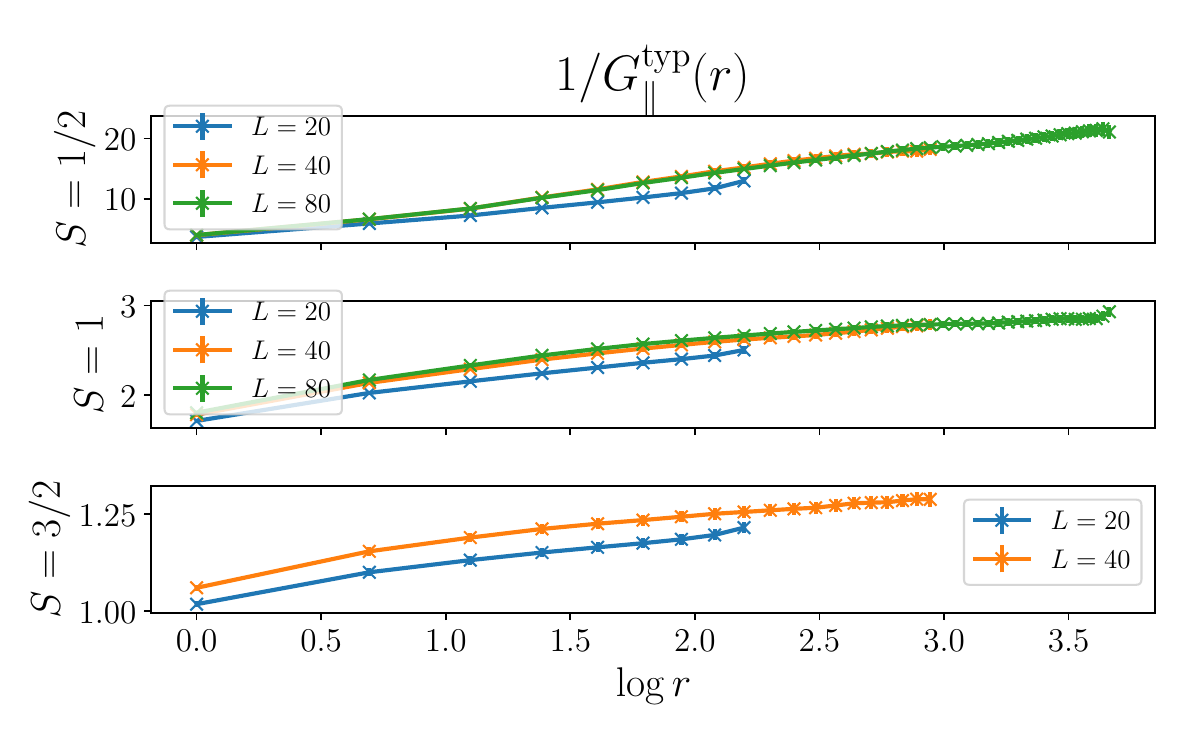}
\caption{{ Plot of $1/G_{\parallel}^{\rm typ}$ vs. $\log r$ for $S=1/2,\,1,\,3/2$. While the trend for $S=1/2$ is harder to interpret, the $S=1,\,3/2$ plots show a clear downward curvature, which we interpret as a result of $G_\parallel$ tending to a nonzero constant at large $r$ (rather than decaying logarithmically \cite{PhysRevB.60.12116}).}}
\label{fig:ZZ-logr}
\end{figure}

{

\section{Scaling of the entanglement entropy}
\label{app:sec:entanglement}

In this section we report numerical calculation of the ground state entanglement entropy within linear spin-wave theory.
Given a bipartition of the chain into two sub-regions $\mathfrak{A}$ and $\mathfrak{B}$, the von Neumann entanglement entropy is given by
\ba
    S_{\rm vN} &= -\Tr_\mathfrak{A} \rho_\mathfrak{A} \log \rho_\mathfrak{A},
    &
    \rho_\mathfrak{A} = \Tr_\mathfrak{B} \ket{\psi} \bra{\psi},
\end{align}
where $\Tr_{\mathfrak{B}}$ denotes the partial trace over  region $\mathfrak{B}$ and in our case $\ket{\psi}$ is the ground state of the system.
We will consider the simplest setting, where the chain has open boundary conditions, $\mathfrak{A}$ contains the sites from $1$ to $r$, and $\mathfrak{B}$ contains the sites from $r+1$ to $L$.

In general,  $S_{\rm vN}(r,L)$ can be expected to have a non-trivial dependence on $r$ and $L$
in a 1D state that supports power-law correlations
{ (states with only area-law entanglement, i.e. $S_{\rm vN}$ of order 1, are usually well-approximated by finite-bond-dimension matrix product states, which cannot have power-law correlations)}. For example, in conformal field theories it is known that~\cite{Calabrese_2009}
\be
\label{eq:cft-entanglement}
S_{\rm vN} = \f{c}{6} \log\left( L \sin\frac{\pi r}{L} \right) + \tilde{b},
\ee
where $c$ is the central charge and $\tilde{b}$ gives a subleading non-universal correction.
{
The present spin wave theory is certainly not conformally invariant (for example, the dynamical exponent is not unity) but}
the  numerical results that follow suggest a qualitatively similar scaling for the average entanglement entropy of our model. We begin by describing how the entanglement entropy can be computed within linear spin-wave theory.

For this purpose, we rely on the observation~\cite{Ingo_Peschel_2003, Peschel_2009} that if $\ket{\psi}$ is a Gaussian state, as in our case, also the reduced density matrix $\rho_\mathfrak{A}$ will be Gaussian. Therefore $\rho_\mathfrak{A}$ and the entanglement entropy can be reconstructed from the quadratic correlators within subsystem $\mathfrak{A}$. Given that to the best of our knowledge the problem of computing the entanglement entropy in a bosonic system with non-zero pairing correlations has not been explicitly discussed in the literature, we report here the details of the method, which is a straightforward generalization of Ref.~\cite{Ingo_Peschel_2003} combined with the canonical transformation of Ref.~\cite{COLPA1978327}.

We define $\vec{\alpha}_{\mathfrak{A}}$ the vector of operators from $\vec{\alpha}$ in~\eqref{eq:H-spin-wave} that act on $\mathfrak{A}$. Then, by gaussianity $\rho_\mathfrak{A}$ must be of the form
\be
    \rho_\mathfrak{A} = Z^{-1} \exp\left( - \vec{\alpha}^\dag_{\mathfrak{A}} \mathcal{A} \vec{\alpha}_{\mathfrak{A}} \right)
\ee
for some normalization $Z$ and matrix $\Gamma$. [Note that more general terms would be incompatible with $\mathrm{U}(1)$ symmetry.] After an appropriate canonical transformation $\vec\gamma = R \vec{\alpha}_{\mathfrak{A}}$, the density matrix can be diagonalized, viz.
\be
    \rho_\mathfrak{A} = Z^{-1} \exp\left( - \sum_k \epsilon_k \gamma_k^\dag \gamma_k \right).
\ee
Here, for $\vec{\gamma}$ to satisfy canonical commutation relations we must have $R J_{0, \mathfrak{A}} R^\dag = J_{0,\mathfrak{A}}$, where $J_{0,\mathfrak{A}}$ is the restriction of the matrix $J_0$ to the sites in $\mathfrak{A}$.

Peschel observed~\cite{Ingo_Peschel_2003} that $\Gamma$ is fully determined by the set of quadratic correlators with support in $\mathfrak{A}$. In our case, the only non-vanishing correlators are contained in the matrix
\be
    C = \braket{\psi| \vec{\alpha}_{\mathfrak{A}} \vec{\alpha}_{\mathfrak{A}}^\dag |\psi}.
\ee
It is then straightforward to see that
\be
    C = R^{-1}
    \left[
        \begin{pmatrix}
            \mathds{1}_{N_{A,\mathfrak{A}}} & 0 \\
            0 & 0
        \end{pmatrix}
        +
        \frac{1}{e^{\epsilon}-1}
    \right]
    (R^{-1})^\dag
\ee
where $N_{A,\mathfrak{A}}$ is the number of $A$ sites in the region $\mathfrak{A}$.
Therefore, the problem of finding the values of $\epsilon$ reduces to the problem of finding the canonical transformation diagonalizing $C$, which can be tackled with the technique discussed in Sec.~\ref{app:sec:spin-wave-theory} and Ref.~\cite{COLPA1978327}.
Finally, the von Neumann entanglement entropy is given by~\cite{Ingo_Peschel_2003}
\be
    S_{\rm vN} = \sum_k \left[ - \ln(1-e^{-\epsilon_k}) + \frac{\epsilon_k}{e^{\epsilon_k} - 1}\right].
\ee

The data obtained in this way is reported in Fig.~\ref{fig:entanglement}. First, we can observe that $S_{\rm vN}$ increases with the lengthscale, as it can be expected in a theory supporting power-law correlators. In the top panel, we see that $S_{\rm vN}^{\rm av}(r=L/2,L)$ grows with system size. Given that the theory is scale invariant, we might expect the growth to be logarithmic, viz. $S^{\rm av}_{\rm vN} = \frac{c_{L/2}}{6} \log L + b'$. A fit of the last three data points indicates that $c_{L/2}\approx 1.5$, although this is probably higher than the asymptotic value  at large $L$ due to finite-size effects.
{We have included the normalization factor $1/6$ to facilitate comparison with familiar critical states, though here $c$ cannot be interpreted as a central charge.}

We might also ask about the dependence on $r$ for a fixed $L$. Given that we are not dealing with the entanglement entropy of a CFT, there is no a priori reason to expect Eq.~\eqref{eq:cft-entanglement} to hold in our model. 
Nonetheless, 
{if we assume 
(i)  that 
$S_{\rm vN}(r=L/2,L)$
and 
$S_{\rm vN}(r,\infty)$
grow logarithmically with the same coefficient, and
(ii)  symmetry under reflection $S_{\rm vN}(r,L) = S_{\rm vN}(L-r,L)$, then the simplest scaling ansatz is 
\be
S_{\rm vN}(r, L) = \frac{c_{L/2}}{6} \ln \lf L\, \mathfrak{s}(r/L) \ri,
\ee
where 
$\mathfrak{s}(r/L) = \lim_{L\to\infty} L^{-1} {\exp (S_{\rm vN}/c_{L/2})}$ can be Fourier expanded as
\be
    \mathfrak{s}(r/L) = \sum_{l=0}^\infty \mathfrak{a}_l \sin\frac{(2l+1)\pi r}{L},
\ee
and we set $\mathfrak{a}_1=1$ without loss of generality.
Here we used assumption (i) to rule out the presence of cosines in the expansion, and (ii) to restrict the coefficient of $\pi r/L$ in the sine to be odd.
This more general form recovers the CFT entanglement scaling~\eqref{eq:cft-entanglement} when $\mathfrak{a}_l=0$ $\forall l>0$.

Given that our numerical data suffers from significant finite-size effects we are unable to perform a better scaling analysis using the scaling forms above. Therefore, we simply show in the bottom panel of Fig.~\ref{fig:entanglement} that $S_{\rm vN}^{\rm av}$ has an approximately linear dependence on $\log\left(L \sin(\pi r / L) \right)$. Performing a linear fit of this kind on the data obtained at $L=1280$ we estimate $c_{L/2}\approx 1.2$ although this is probably smaller than the asymptotic value at $L=\infty$ given that the estimate for $c_{L/2}$ increases slightly for increasing system sizes.

\begin{figure}
    \centering
    \includegraphics[width=\linewidth]{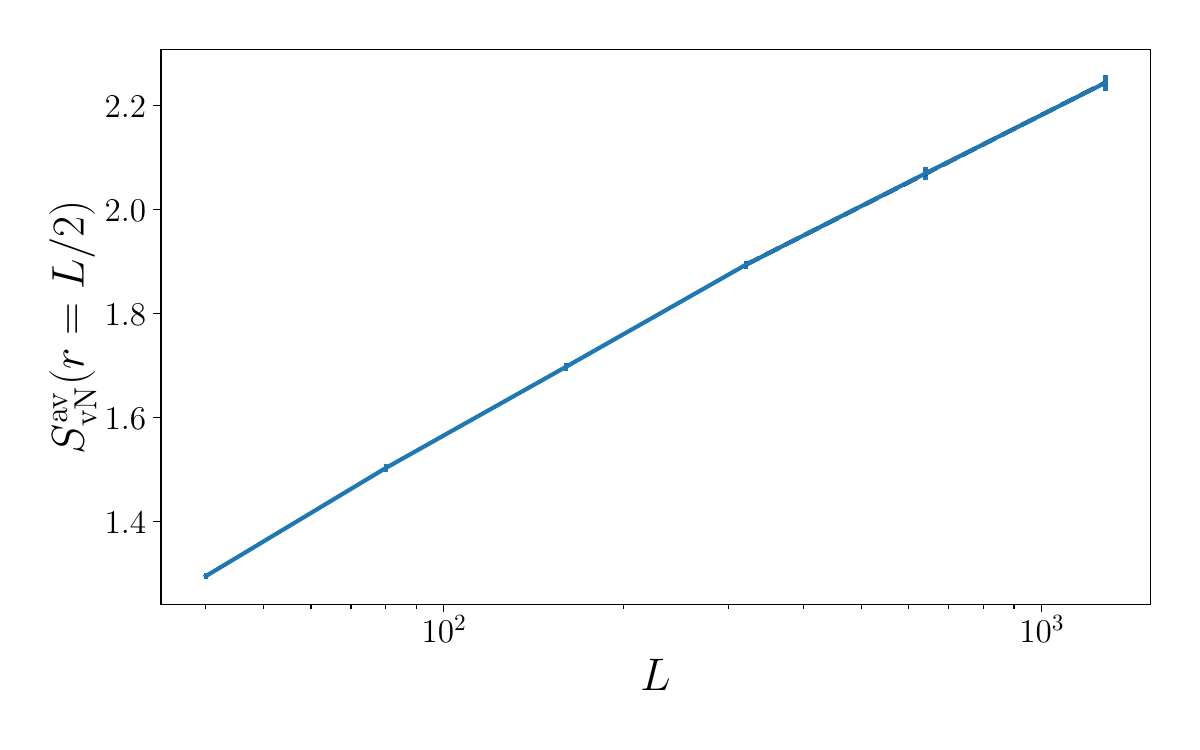}\\
    \includegraphics[width=\linewidth]{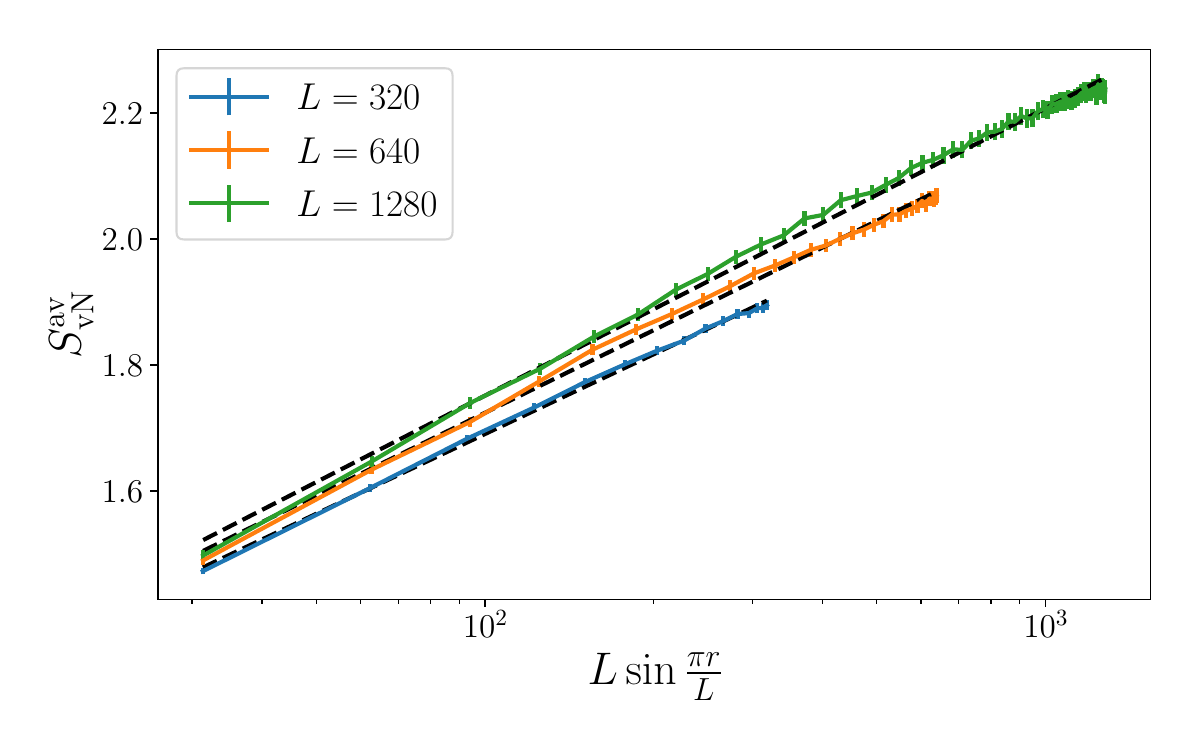}
    \caption{{ Disorder-averaged von Neumann entanglement entropy $S_{\rm vN}(r,L)$ for a cut at site $r$ in a system with open boundary conditions.  Calculations are performed within linear spin-wave theory in the presence of a uniform field of magnitude $10^-7 J$. Top panel: scaling of $S_{\rm vN}(L/2,L)$ as a function of $L$. Bottom panel: $S_{\rm vN}(r,L)$ as a function of $r$. Different curves correspond to different values of $L$. The $x$-axis is chosen in such a way that for a CFT the data would fall on a single straight line (note the logarithmic scale). Black dashed lines are obtained through linear fits.}}
    \label{fig:entanglement}
\end{figure}

}
}

\section{Further DMRG data: correlators}
\label{sec:further-DMRG}

In this section we report further data pertaining to correlation functions for different values of $S$.
The longitudinal correlators (typical and average) are 
\ba\label{eq:applong}
G_{\parallel}^{\rm av}(r) &= \f{1}{S}\overline{\langle N_{j}^x N_{j+r}^x\rangle},\\
\log G_{\parallel}^{\rm typ}(r) &= \overline{\log \left[ S^{-1}\langle N_{j}^x N_{j+r}^x\rangle \right] },
\end{align}
and the transverse correlators are 
\ba\label{eq:apptransverse}
G_{N}^{\rm av}(r) &= \f{1}{2S}\overline{\langle \vec{N}_{j}^\perp .\vec{N}_{j+r}^\perp \rangle},\\
\log G_{N}^{\rm typ}(r) &= \overline{\log \left[ (2S)^{-1} \langle \vec{N}_{j}^\perp .\vec{N}_{j+r}^\perp\rangle \right] }.
\end{align}
The overline indicates averaging over disorder and also over the coordinate $j$ for a given sample.
For a given value of $r$, we average over $j$ values in the range ${[L/4+1, 3L/4-r]}$.
The normalization $1/2S$ in (\ref{eq:apptransverse}) 
is chosen to match the convention we used in discussing spin wave theory. 
The normalization $1/S$ in (\ref{eq:applong}) is then chosen so that,
in the absence of any breaking of $\mathrm{SO}(3)$ symmetry, the longitudinal and transverse correlators would be equal.
(With this normalization, the value  of $G_{\parallel}$ for a perfectly ordered product state would be $S$.)

The three panels of Fig.~\ref{fig:comparecorrelators}
compare the longitudinal and
transverse correlators (typical values) for the cases of ${S=1/2}$, ${S=1}$ and ${S=3/2}$ respectively.
As noted in the main text, there  is a strong asymmetry between longitudinal and transverse correlators, and the longitudinal correlators seem consistent with convergence to a nonzero value at large distances, 
i.e. with long range order.

It was proposed in Ref.~\onlinecite{PhysRevB.60.12116}, on the basis of a numerical strong-disorder renormalization group scheme for a chain with a box distribution of couplings in ${[-J_0, J_0]}$, that the two-point function ${\< \vec{N}_j \cdot \vec{N}_k\>}$ decayed as ${a/\ln (r/r_0)}$. 
This scaling would of course be incompatible with the long range order proposed here for the ${\pm \mathcal{J}}$ chain.
Since the logarithmic decay is very slow, we cannot rule it out on numerical grounds. However, spontaneous symmetry breaking with a nonzero asymptote for $G_\parallel$ seems the simplest explanation for Fig.~\ref{fig:comparecorrelators}.
In Fig.~\ref{fig:ZZ-logr} we show $1/G_\parallel$ as a function of $\log r$ (the prediction of 
Ref.~\onlinecite{PhysRevB.60.12116} would be a straight line on these plots).

Spin wave theory predicts that the transverse correlator should decay as $r^{-1/2}$. 
On the available sizes, the transverse correlators 
decay faster than this prediction, with the deviation being larger for smaller $S$
(see the trend lines in Fig.~\ref{fig:comparecorrelators}).
However, it seems plausible to us that this is due to finite size effects (because $r$ is not large enough and/or because $r/L$ is not small enough).
For ${S=1/2}$ the deviation is surprisingly large (the effective exponent is closer to $-1$ on these scales), 
but the behavior of the longitudinal component suggests that finite size effects are very large on these scales
(the longitudinal correlator is varying significantly over these lengthscales, 
whereas if the chain is indeed long-range ordered this correlator will be approximately constant at very large distances).

\section{DMRG details and convergence}
\label{sec:DMRG-convergence}

In this appendix we present further data related to the convergence of the DMRG simulations.
We begin, however, by describing the parameters of the algorithm employed.

Each ground state search is performed as follows. We start by optimizing a MPS with maximum bond dimension $\chi_0$
--- throughout all DMRG sweeps we neglect singular values below a fixed threshold of $10^{-12}$ ---
and perform DMRG sweeps with a fixed ``noise''~\cite{PhysRevB.72.180403,SCHOLLWOCK201196} magnitude --- in practice we find that a magnitude in the range $[10^{-9},10^{-8}]$ works well. We stop the DMRG when during a sweep the relative energy change is lower than $10^{-12}$ and, in any case, not before some fixed number $N_{\rm min}$ of sweeps have been performed. (In practice we keep $20<N_{\rm min}<100$ according to how fast observables converge.) We then optimize further the state through the same procedure, but without using noise. After recording the observables of interest, we  increase the maximum allowed bond dimension to $\chi_1$ and iterate the same procedure. In this way we obtain results for a sequence of maximum bond dimensions $[\chi_0,\chi_1,\chi_2,\dots]$. We checked that the data obtained is converged in the sense that
the final iteration step with the largest $\chi$
produces a relative change of less than $10^{-3}$ in the  order parameter and energy gap.
For the cases we consider, we find that a maximum bond dimension of $\chi=800$ or $1200$, depending on the system size, is sufficient.

Next we discuss the reliability of the data obtained through DMRG. Note that this is not a trivial matter in disordered system, and the standard check of convergence of observables as a function of the bond dimension, as recorded above, might not be sufficient.
For example, DMRG is known to have difficulty capturing the correlations of infinite-randomness fixed points (IRFPs) (see e.g. Ref.~\onlinecite{IGLOI2005277} for a review). In IRFPs energy scales become exponentially small at large lengthscales, therefore DMRG tends to neglect large scale correlations that give only an exponentially small energy gain. However, in the model under consideration we expect energy scales to have only a power law dependence on the length scale, and therefore DMRG could be expected to give accurate results.

As a numerically accessible indicator of convergence we use the ratio
\begin{equation}
\label{eq:epsilon-def}
\epsilon = \frac{\braket{\psi|H^2|\psi} - \braket{\psi|H|\psi}^2}{\Delta E^2},
\end{equation}
where $\Delta E$ is the energy gap between the sectors with spin $S^x=S|n_A-n_B|$ and $S^x=S(|n_A-n_B|+1)$ reported in the main text. We use $\epsilon$ as a proxy for a similar quantity
\begin{equation}
\tilde{\epsilon} = \frac{\braket{\psi|H^2|\psi} - \braket{\psi|H|\psi}^2}{\widetilde{\Delta E}^2},
\end{equation}
with $\widetilde{\Delta E}$ denoting the gap within the sector ${S^x=S|n_A-n_B|}$.
While ${\Delta E \geq \widetilde{\Delta E}}$, 
effectively we assume that the rough magnitudes of these two gaps are comparable, writing $\tilde\epsilon \approx\epsilon$.
If 
\begin{equation}\label{eq:definevarepsilon}
\ket{\psi} = \sqrt{1-\varepsilon} \ket{GS} + \sqrt{\varepsilon} \ket{\phi},
\end{equation}
where $\ket{GS}$ is the exact ground state, $\ket{\phi}$ a normalized state orthogonal to $\ket{GS}$, and $\varepsilon$ a positive parameter that determines how good the convergence to the ground state is, then 
\be
\tilde \epsilon \leq \varepsilon (1-\varepsilon).
\ee
In principle it is possible to achieve a small $\tilde\epsilon$ with a state that has a very small overlap with the ground state (e.g. if the DMRG was to converge to an excited eigenstate, giving ${\varepsilon=1}$), but this seems very unlikely. Therefore we assume that, when $\tilde \epsilon$ is small, the error is bounded by
\be
\varepsilon \lesssim \tilde \epsilon \approx \epsilon.
\ee

In Fig.~\ref{fig:variance-hist}, we report the histogram of $\epsilon$ for a few representative system sizes and spin values $S$. At a qualitative level, the fact that $\epsilon\ll 1$ for most disorder realizations indicates that in most cases DMRG produces a state whose overlap with the exact ground state is close to 1.
For large sizes, there are a few  rare samples which do not converge close to the ground state. We believe that this is because the distribution of the energy gap is relatively broad, giving some samples with a very small gap which are challenging for DMRG. As a result of these rare samples we must carefully bound the possible systematic error from DMRG.

\begin{figure}
\centering
\includegraphics[width=\linewidth]{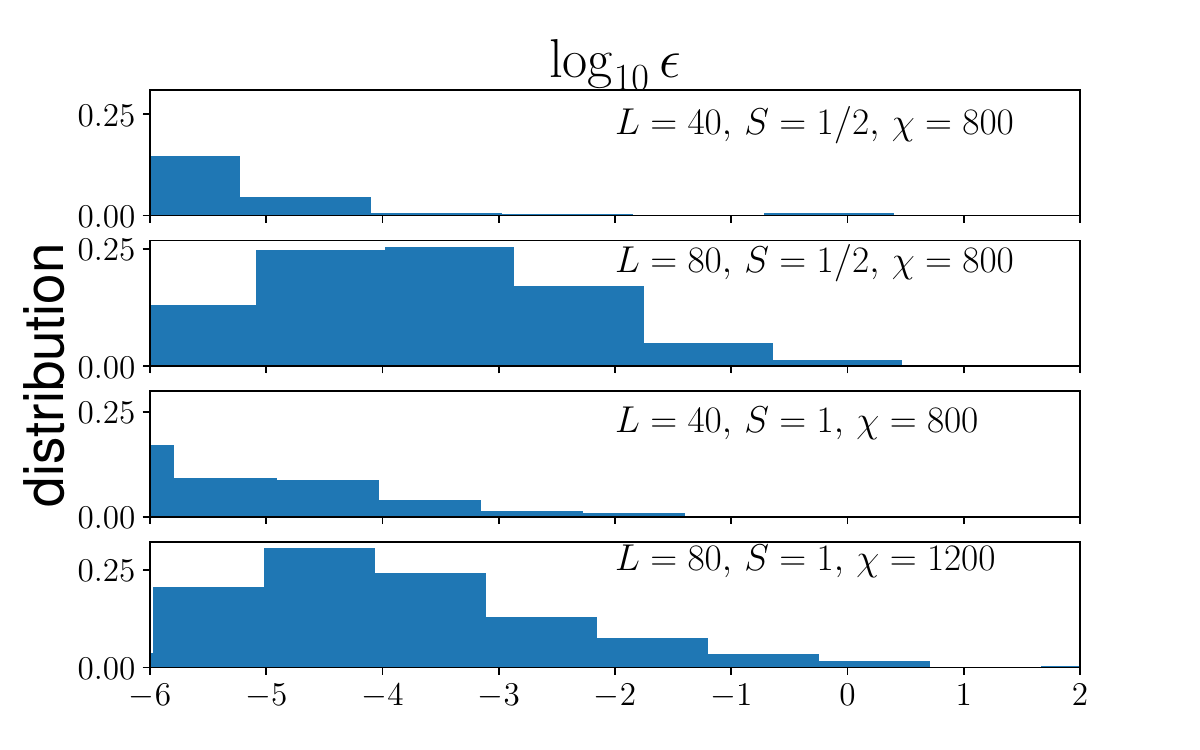}
\caption{Normalized histogram of $\epsilon$, as defined in~\eqref{eq:epsilon-def}. For graphical convenience, only data with $\epsilon>10^{-6}$ is shown. For most disorder realizations $\epsilon \ll 1$, signalling a good DMRG convergence.}
\label{fig:variance-hist}
\end{figure}

We now provide quantitative bounds on any systematic errors on the data obtained through DMRG.
We consider the energy gap and the order parameter separately, since the former is more straightforward.
In both cases the analysis confirms that the possible systematic error is small. Our  estimate for the maximum possible systematic error in the order parameter was marked on 
Fig.~2 for the large sizes.

For a numerically accessible estimate of the maximum order of magnitude of the error in the determination of the ground state energy, we use the energy variance of the state:
\be\label{eq:varianceestimate}
|\<\psi| H-E_0 | \psi \>| \lesssim 
\sqrt{ \<  \psi| H^2 | \psi \>  - 
\<  \psi| H| \psi \>^2}.
\ee
The RHS is  ${\sqrt{\epsilon} \Delta E}$ in the notation of Eq.~\ref{eq:epsilon-def}.
We will use this as an order-of-magnitude upper bound for all samples, but we note that for the well-converged samples (which are the majority --- these are the ones with small $\varepsilon$, see  Eq.~\ref{eq:definevarepsilon}) it can be related to a precise bound.
First note that
\be\label{eq:energybound1}
\< \psi| H-E_0 |\psi\>^2 \leq
\< \psi| (H-E_0)^2 |\psi\>.
\ee
The RHS is related to the quantities $\epsilon$ and $\varepsilon$ defined at the beginning of this Appendix by
\ba\label{eq:energybound2}
\epsilon \Delta E^2 
&\equiv 
\< \psi | H^2 | \psi\> 
- \< \psi | H | \psi\>^2 \\ \notag
& = 
\< \psi | (H-E_0)^2 | \psi\> 
- \< \psi | H-E_0 | \psi\>^2
\\ \notag
& = \varepsilon \< \phi | (H-E_0)^2 | \phi\> 
- \varepsilon^2 \< \phi | H-E_0 | \phi\>^2
\\  \notag
& \geq 
\varepsilon \< \phi | (H-E_0)^2 | \phi\> 
- \varepsilon^2 \< \phi | (H-E_0)^2 | \phi\>
\\ \notag
& = (1-\varepsilon) \<\psi| (H-E_0)^2 |\psi\>,
\end{align}
i.e. from (\ref{eq:energybound1},~\ref{eq:energybound2})
\be
|\< \psi| H-E_0 |\psi\>|  \leq \f{\sqrt{\epsilon}\,\Delta E}{\sqrt{1-\varepsilon}}
\ee
which is close to (\ref{eq:varianceestimate}) when $\varepsilon$ is small.

Since we compute the gap $\Delta E$ as the energy difference between two ground states in two different $S^x_{\rm tot}$ sectors,
the error in the excited state energy can be computed in the same way as in Eq.~\ref{eq:varianceestimate}.
We  estimate the maximum error in the energy \textit{gap} as the sum of these maximum errors for the ground and excited states, for a given sample.
Finally this gives an error bound for the sample-averaged gap:
\begin{equation}
\label{eq:gap-error-bound}
\left|\Delta E^{\rm av} - \Delta E_{(\textrm{exact})}^{\rm av}\right|\leq
\overline{\Delta E \left(\sqrt{\epsilon} +  \sqrt{\epsilon^{(\textrm{exc})}}\right)}.
\end{equation}
We report the value of this bound in Fig.~\ref{fig:gap-error-bounds}.
In all cases the error bound is $\lesssim 2\%$, and much smaller than the associated statistical error from averaging over samples.

\begin{figure}
\centering
\includegraphics[width=\linewidth]{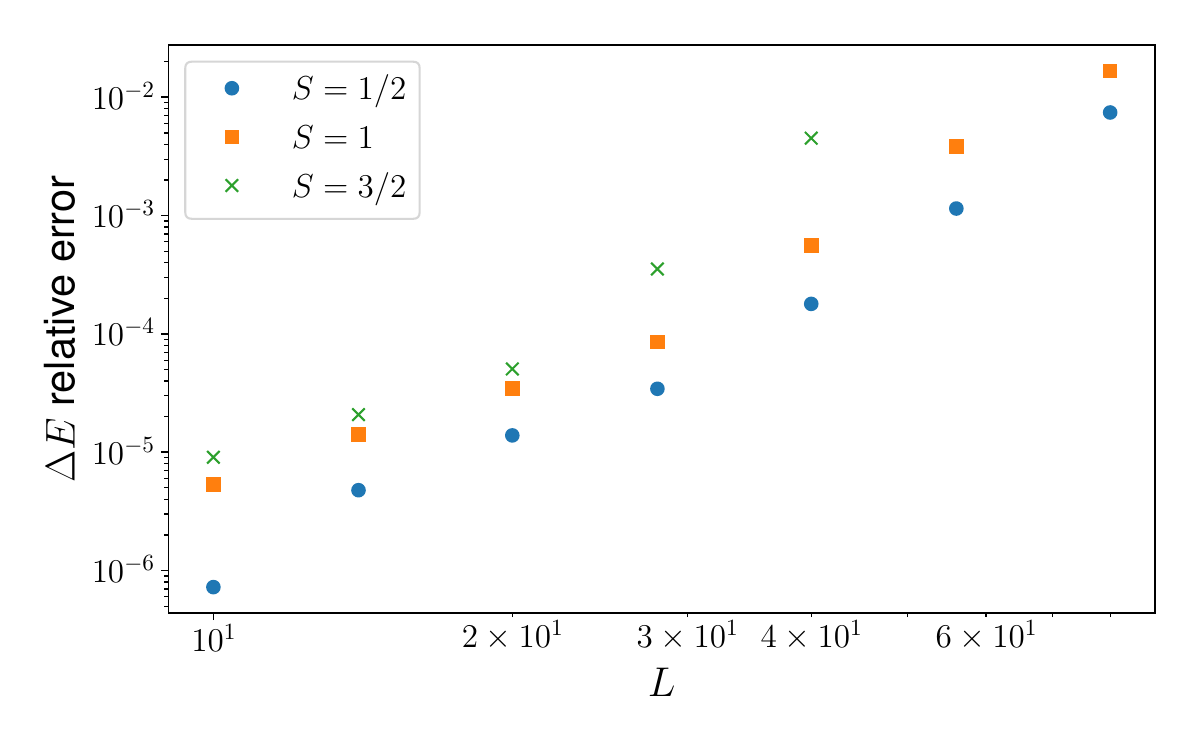}
\caption{Bound on the systematic error on the average gap $\Delta E^{\rm av}$, as obtained through Eq.~\ref{eq:gap-error-bound}. In all cases the bound on the systematic error is $\lesssim 2\%$ and much smaller than the associated statistical error.}
\label{fig:gap-error-bounds}
\end{figure}

Next, we turn to the order parameter.
Our aim is to quantify the worst-case effect that rare samples which are badly converged could have on the estimate.

For this purpose, we set a cutoff value ${\epsilon_*=0.1}$, dividing the samples into two cohorts:

\begin{figure*}[t]
\centering
\includegraphics[width=0.47\linewidth]{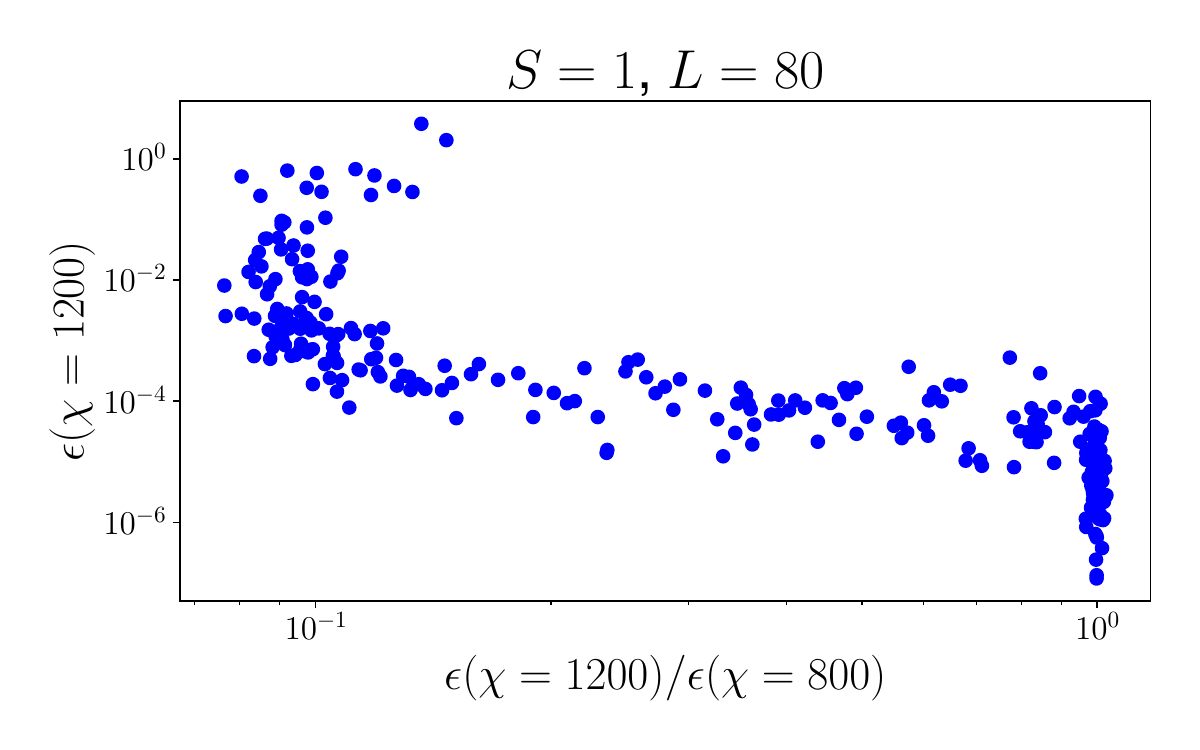}
\includegraphics[width=0.47\linewidth]{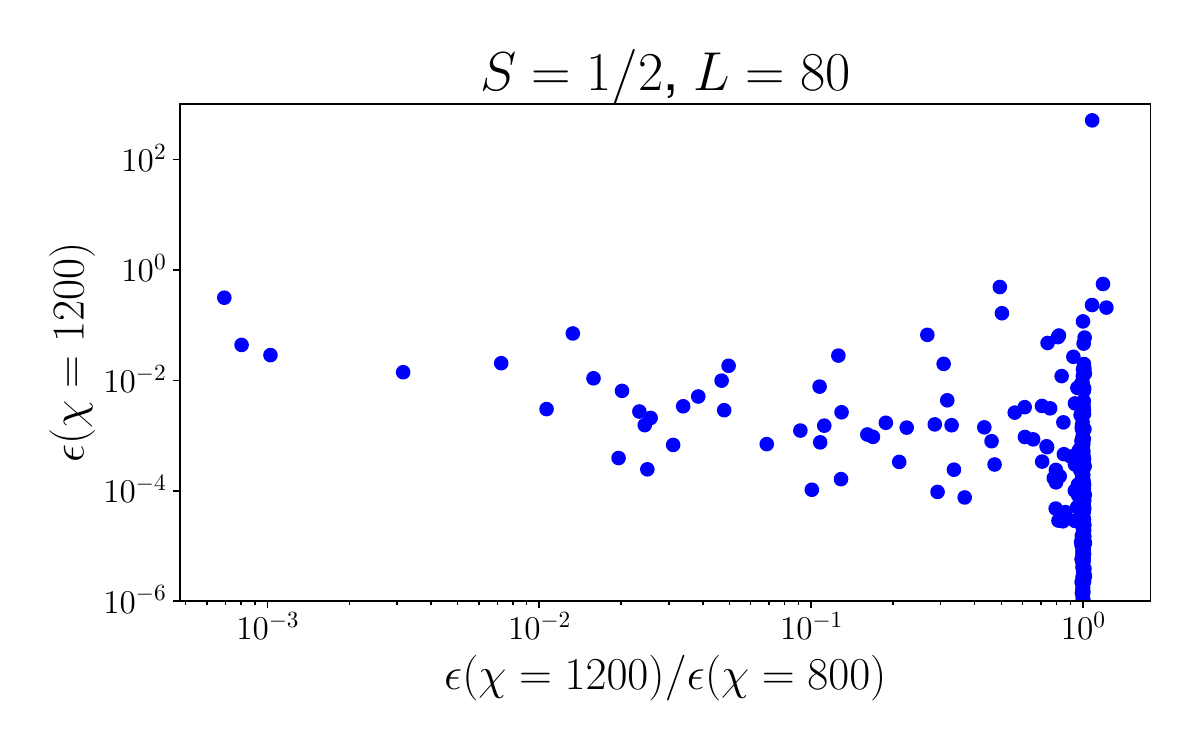}
\caption{ 
Left: Scatter plot of the $\epsilon$ achieved at the maximum bond dimension ($y$-axis), and $\epsilon$ reduction upon the last bond dimension increase, for  $S=1$ chains with $L=80$. Each point corresponds to a different disorder realization. We see that either $\epsilon$ is already very small or it is significantly reduced during the last $\chi$ increment. 
Right: A similar scatter plot  for $S=1/2$.
For $S=1/2$, in some disorder realizations, $\epsilon$ saturates to a large value.
}
\label{fig:eps-decrease-S-1}
\end{figure*}

For the cohort of samples with $\epsilon<\epsilon_*$, DMRG has found a state that is close to the ground state. 
For this cohort we believe that a standard error analysis, from the variation of the order parameter as a function of the bond dimension, gives a reasonable estimate for the order of magnitude of the systematic error, and shows that it is smaller than the statistical error.
[For spin-1, the evidence for this statement  is that the  error $\epsilon$  for  these samples decreases significantly with bond dimension (except for samples that have already achieved a very small error $\epsilon \lesssim 10^{-4}$)
without significant change in the resulting order parameter estimate (see Figs.~\ref{fig:eps-decrease-S-1}
below). 
For the spin-1/2 case, there are samples with  $\epsilon<\epsilon_*$ whose error $\epsilon$ did not improve significantly with the bond dimension. However, we reran a fraction of samples with more Davidson steps. This did achieve significantly better convergence for almost all of these samples, but again the change in the estimate of the order parameter was negligible compared to the statistical error.]

For the (small) cohort of ``bad'' samples with $\epsilon>\epsilon_*$,  some of which are not converging with bond dimension,
it is not clear how to estimate the error in a standard way.
Therefore for these samples we make the most conservative possible  error estimate by comparing with the result that would be obtained if we set  ${\< GS | N_j^x | GS \> =0}$ for these samples. Since the  quantity on the LHS is necessarily non-negative (a consequence of the Marshall sign rule), this gives a lower bound on the expectation value for these samples.
The results of this procedure were shown for some of the largest system sizes in Fig.~2 (main panel) of the main text.

Finally, for the study of correlation functions, we proceed along similar lines. However, the fact that  we consider typical values  for the correlator, which are obtained by averaging $\log \langle \vec{N}_j^\perp \cdot \vec{N}_{j+r}^\perp \rangle$,
means we cannot simply use the bound $\langle \vec{N}_j^\perp \cdot \vec{N}_{j+r}^\perp \rangle \geq 0$ 
(the logarithm of 0 would give a divergence).
We assume that a reasonably conservative error estimate for $G_N^{\rm typ}$ 
is given by asking how the final result would change if,
for the cohort of poorly-converged samples with $\epsilon>\epsilon_*$, 
the DMRG estimates of $\langle \vec{N}_j^\perp \cdot \vec{N}_{j+r}^\perp \rangle$ and $\langle N^x_j N_{j+r}^x \rangle$
calculation  were wrong by a factor of $2$. In all the figures reported 
we find that this has negligible effects on $G_N^{\rm typ}$ and $G_{\parallel}^{\rm typ}$. For example, this would produce a maximum relative displacement of $3\%$ on $G_N^{\rm typ}$ for $S=1$ and $L=80$.
Thus, we have not reported these error estimates in Fig.~3 of the main text and in the figures of App.~\ref{sec:further-DMRG}.

Finally we give some more data on convergence.
Fig.~\ref{fig:eps-decrease-S-1} 
shows scatter plots of the $\epsilon$ values (see Eq.~\ref{eq:epsilon-def}) achieved by the various samples at the maximum bond dimension, together with the $\epsilon$ reduction obtained in the last bond dimension increase.

}

\end{appendix}

\end{document}